\documentclass[secnumarabic,aps,prd,floatfix,nofootinbib,showkeys,twocolumn,showpacs,10pt,superscriptaddress] 
{revtex4-1}

\usepackage{times}
\usepackage{amsfonts}
\usepackage{amsmath}
\usepackage{amssymb}
\usepackage{natbib}
\usepackage{graphicx}
\usepackage{enumitem}
\usepackage{xcolor}
\usepackage[colorlinks=true,linkcolor=red, citecolor=blue]{hyperref}
\usepackage[outdir=./]{epstopdf}
\usepackage[normalem]{ulem}
\usepackage{amsmath}

\def\aap{A\&A}

\def\apj{ApJ}
\def\apjl{ApJL}
\def\apjs{ApJS}

\def\jcap{JCAP}
\def\mnras{MNRAS}
\def\na{New Astronomy}

\def\physrep{Physics Reports}

\begin{document}
\definecolor{orange}{rgb}{0.9,0.45,0}
\def\CovDev{D}
\def\Res{{\mathcal R}}
\def\Gammaflat{\hat \Gamma}
\def\metricflat{\hat \gamma}
\def\Dflat{\hat {\mathcal D}}
\def\part_n{\partial_\perp}
%
\def\Lie{\mathcal{L}}
\def\A{\mathcal{X}}
\def\Aphi{\A_{\phi}}
\def\hAphi{\hat{\A}_{\phi}}
\def\E{\mathcal{E}}
\def\Ham{\mathcal{H}}
\def\M{\mathcal{M}}
\def\R{\mathcal{R}}
\def\p{\partial}
\def\hg{\hat{\gamma}}
\def\hA{\hat{A}}
\def\hD{\hat{D}}
\def\hE{\hat{E}}
\def\hR{\hat{R}}
\def\hcA{\hat{\mathcal{A}}}
\def\hDelt{\hat{\triangle}}
\def\na{\nabla}
\def\dif{{\rm{d}}}
\def\non{\nonumber}
\newcommand{\erf}{\textrm{erf}}
\newcommand{\saeed}[1]{\textcolor{blue}{SF: #1}} 
%
\renewcommand{\t}{\times}
\long\def\symbolfootnote[#1]#2{\begingroup%
\def\thefootnote{\fnsymbol{footnote}}\footnote[#1]{#2}\endgroup}
\title{Toward More Realistic Mass Functions for Ultradense Dark Matter Halos} 

\author{Saeed Fakhry}
\email{s\_fakhry@sbu.ac.ir}
\affiliation{Department of Physics, Shahid Beheshti University, 1983969411, Tehran, Iran }

\author{Marzieh Farhang}
\email{m\_farhang@sbu.ac.ir}
\affiliation{Department of Physics, Shahid Beheshti University, 1983969411, Tehran, Iran }

\author{Antonino Del Popolo} 
\email{antonino.delpopolo@unict.it}
\affiliation{Dipartimento di Fisica e Astronomia, University of Catania, Viale Andrea Doria 6, 95125 Catania, Italy}
\affiliation{Institute of Astronomy, Russian Academy of Sciences, Pyatnitskaya str. 48, 119017 Moscow, Russia}
\affiliation{Institute of Astronomy and National Astronomical Observatory, Bulgarian Academy of Sciences, 72, Tsarigradsko Shose Blvd., 1784 Sofia, Bulgaria}
\date{\today}

\begin{abstract} 
\noindent
Ultradense dark matter halos (UDMHs) are high concentrations of dark matter, assumed to have formed deep in the radiation-dominated era from amplified primordial perturbations. In this work we improve the previous works for the calculation of UDMH abundance by elaborating on the formation process of these halos by including various physical and geometrical modifications in the analysis. In particular, we investigate the impact of angular momentum, dynamical friction and triaxial collapse on the predicted mass functions for UDMHs. We perform the calculations for four primordial power spectra with different amplified features that allow for primordial black hole and UDHM formation in wide and narrow mass ranges. We also apply this analysis in the context of two possible scenarios for dark matter: the single-component and the multi-component. We find that the abundance of UDMHs is prominently enhanced in the presence of these more realistic mass functions. 
\end{abstract}

\keywords{Dark Matter --- Halo Mass Function --- Primordial Black Hole --- Power Spectrum --- Gravitational Collapse}

\maketitle
\vspace{0.8cm}

\section{Introduction} 
In cosmic evolution, the structure formation process is expected to follow a bottom-up path. Initially, smaller bound objects form from the gravitational collapse of random density perturbations. Such structures subsequently merge to form larger structures during the evolution of the Universe. The simple scenario is based on the assumption that density fluctuations arise from inflationary-induced primordial perturbations, described by an almost scale-invariant spectrum, see, e.g., \citep{1972MNRAS.160P...1Z, 1982PhLB..117..175S, 1982PhRvL..49.1110G, 1984PThPS..78....1K, 1992PhR...215..203M}. Most of these perturbations have small amplitudes ($\delta\sim 10^{-5}$) when they enter the cosmic horizon that does not collapse until long after the matter-radiation equality. However, these assumptions are not necessarily universal. For example, inflation models often predict deviations from scale invariance or Gaussian statistics, see, e.g., \cite{1992JETPL..55..489S, 2000PhRvD..62d3508C, 2001PhRvD..63l3501M, 2008PhRvD..77b3514J, 2010AdAst2010E..76B, 2010CQGra..27l4010K, 2011JCAP...05..024A}. Additionally, more complex cosmological scenarios, including e.g., phase transitions and late-time dynamical scalar fields, may lead to additional fluctuations at certain scales, see, e.g., \citep{1997NuPhB.503..405A, 2009JCAP...02..014A, 2011PhRvD..84h3503E, 2018JCAP...07..007D}. As a result, structures on these scales can collapse significantly earlier than expected in the standard scenario, i.e., before the matter-radiation equality.

Primordial black holes (PBHs) are one of the most widely studied consequences of such early gravitational collapse. PBHs are macroscopic candidates for dark matter, which are believed to form from small-scale, large-amplitude primordial density fluctuations ($\delta>0.3$) during the early stages of the Universe, see, e.g., \citep{2021FrASS...8...87V, 2021JPhG...48d3001G, 2021arXiv211002821C, 2023JCAP...05..054K}. PBHs can have significant gravitational interactions that affect various cosmological observables. They can act as gravitational lenses \citep{2021PhRvD.104h3515W, 2023JCAP...03..043C}, potentially influence the formation of large-scale structures, and also contribute to temperature fluctuations in the cosmic microwave background \citep{2000ApL&C..37..315S}. PBHs have also been proposed to explain gravitational waves resulting from black hole mergers \citep{2016PhRvL.116t1301B, 2016PhRvL.117f1101S} and act as initial seeds for the formation of supermassive black holes \citep{2018MNRAS.478.3756C}.

The constraints on the abundance of PBHs heavily depend on the model for their mass distribution \citep{2021RPPh...84k6902C}. Various studies suggest that PBH mergers within a narrow mass range during the late-time Universe could align with the detected gravitational waves arising from the mergers of stellar-mass black holes, as detected by LIGO-Virgo-Kagra (LVK), especially if they proactively contribute to the dark matter content \citep{2016PhRvL.116t1301B, 2021PhRvD.103l3014F, 2022PhRvD.105d3525F, 2022ApJ...941...36F, 2023PDU....4101244F, 2023PhRvD.107f3507F, 2023ApJ...947...46F, 2023arXiv230811049F, 2024arXiv240115171F}. However, other studies indicate that the scenario based on the merger of PBHs in the early Universe cannot contribute to the merger events recorded by LVK detectors unless their contribution to dark matter is subpercent \citep{2020PhRvD.102l3524H, 2021JCAP...03..068H, 2022PhLB..82937040C, 2022PhRvD.106l3526F}. Besides a narrow mass range for PBHs as the single dark matter component, one can consider PBHs with extended mass distributions and/or scenarios where dark matter consists of various components with distinct masses, interactions, and behaviors, such as PBHs and microscopic particles. The former scenario can address challenges such as initial seeds of supermassive black holes \citep{2018MNRAS.478.3756C} and the cored density profiles of dwarf galactic halos \citep{2020MNRAS.492.5218B}. Also, this scenario, with its broad mass range, can potentially provide a more comprehensive explanation for the detection of gravitational wave signals from merging black holes, see, e.g., \citep{2018CQGra..35f3001S, 2022JCAP...08..006M, 2022PDU....3801115B}.

In both scenarios, dark matter is expected to be present during the formation of PBHs where significant primordial density fluctuations play a crucial role. However, if the primordial density fluctuations are small compared to the collapse threshold, the conditions for PBH formation cannot be fulfilled, and only extremely dense clouds of radiation and dark matter remain. This mechanism requires density contrasts of order $\delta\gtrsim 10^{-3}$ to proceed, which can be much more accessible than those required for PBH formation. Such processes can potentially lead to the formation of ultradense dark matter halos (UDMHs) \citep{2009PhRvL.103u1301S, 2009ApJ...707..979R}. The mass of UDMHs is directly related to the mass of the horizon at the moment when density fluctuations enter it. In simpler terms, the wavelength of these fluctuations as they cross the horizon significantly influences the resulting UDMH mass. Smaller-scale perturbations, with a given amplitude, have an advantage: they enter the horizon earlier, allowing more time for the overdensity to grow. Unlike PBHs, UDMHs lack an event horizon, leading to radiation instability within them. Consequently, radiation escapes from the ultradense regions associated with UDMHs, leaving dark matter as the sole component responsible for their formation \citep{2012PhRvD..85l5027B}. UDMHs exhibit increased sensitivity to small-scale perturbations. Consequently, they can serve as more effective tools for investigating the primordial power spectrum at small scales when compared to PBHs, see, e.g., \citep{2002PhRvD..66j3508T, 2006ApJS..163...80M, 2010MNRAS.404...60R, 2010PhRvD..82h3527J}.

The formation of dark matter halos has been studied through numerical simulations and analytical approaches, specifically targeting those that developed during the matter-dominated era, see, e.g., \citep{2017PhRvD..96l3519G, 2018PhRvD..98f3527D, 2022MNRAS.517L..46W, 2023PDU....4101259D}. Several investigations have also been carried out to specifically examine the emergence of UDMHs during the radiation-dominated era, see, e.g., \citep{1994PhRvD..50..769K, 2002JETP...94....1D, 2010PhRvD..81j3529B, 2013JCAP...11..059B, 2019PhRvD..99l3530N}. Motivated by simulations, \citep{2019PhRvD.100j3010B, 2023MNRAS.520.4370D} introduced an analytical framework to explore the formation and evolution of UDMHs during the radiation-dominated era. This requires definite models for halo mass functions that characterize the distribution of mass among dark matter halos. A large amount of research has been devoted to identifying mass functions that accurately describe the simulations and data of galactic halos, see, e.g., \citep{1974ApJ...187..425P, 2001MNRAS.323....1S, 2002MNRAS.329...61S, 2003MNRAS.346..565R, 2005MNRAS.357...82R, 2006ApJ...646..881W, 2006ApJ...637...12D, 2008ApJ...688..709T, 2017JCAP...03..032D}. In particular, \cite{2023MNRAS.520.4370D} used Monte Carlo simulation to get the distribution of first barrier crossings for a Gaussian random walk which led to the Press-Schechter (PS) mass function \citep{1974ApJ...187..425P}. However, the PS mass function has limitations in accurately describing the mass distribution of dark matter halos, particularly at high redshifts \citep{2017JCAP...03..032D}. The discrepancy can be associated with certain physical effects ignored in the PS formalism which play crucial roles in forecasting halo abundance. 

In this work, we propose to study the formation and evolution scenarios of UDMHs by taking into account various physical and geometrical modifications. In our analysis, we explore two distinct scenarios related to dark matter. The first scenario is a single-component (SC) model, where dark matter exclusively consists of PBHs with various masses. In this model, UDMHs must have formed through the clustering of smaller PBHs. In contrast, the second scenario involves a multi-component (MC) system, where dark matter comprises a mixture of particles and PBHs. The overall structure of this work is outlined as follows. In Section\,\ref{sec:ii} we discuss the formation and evolution of UDMHs during the radiation-dominated era and calculate their abundance with more realistic mass functions and well-motivated primordial power spectra. In Section\,\ref{sec:iii} we present and discuss the results. We conclude in Section\,\ref{sec:iv}.
\section{Ultradense Dark Matter Halos} \label{sec:ii}
In this section we investigate the impact of certain physically-motivated modifications to the formation of UDMHs on their mass functions. We also introduce several primordial power spectra for curvature perturbations, either motivated by various scenarios of the early Universe or considered as possible phenomenological extensions to the scale-invariant spectrum. These spectra will later be used as case studies required to calculate the UDMH mass functions. 
\subsection{Halo formation}
The growth of linear density perturbations of dark matter in the radiation-dominated era, seeded by primordial curvature perturbations $\zeta$, is described by
\begin{eqnarray}\label{eq1}
\delta(k, a) = \alpha \zeta(k) \log\left(\beta\frac{a}{a_{\rm H}}\right) \hspace{1.2cm} \nonumber \\
=\alpha \zeta(k)\log\left(\sqrt{2}\beta\frac{k}{k_{\rm eq}}\frac{a}{a_{\rm eq}}\right),
\end{eqnarray}
where $a\gg a_{\rm H}$, with $a_{\rm H}$ being the scale factor at horizon entry, $k_{\rm eq}\simeq 0.01\,{\rm Mpc^{-1}}$ and $a_{\rm eq} \simeq 3\times 10^{-4}$ are respectively the horizon scale and the scale factor at matter-radiation equality. Moreover we have $\alpha = 6.4$, and $\beta = 0.47$ from numerical suggestions \citep{1996ApJ...471..542H}. Also, horizon crossing occurs at 
\begin{eqnarray}
\frac{a_{\rm H}}{a_{\rm eq}}=\frac{1+\sqrt{1+8(k/k_{\rm eq})^{2}}}{4(k/k_{\rm eq})^{2}}
\simeq \frac{\sqrt{2}}{2}\frac{k_{\rm eq}}{k}, \hspace*{0.4cm} k\gg k_{\rm eq}.
\end{eqnarray}

We assume that at horizon crossing non-relativistic dark matter was decoupled from radiation. The trajectories of individual dark matter particles in response to the curvature kick can be modeled by allowing for independent drift along each axis of an ellipsoid. The parameters of eccentricity, $e$, and prolateness, $p$, characterize the initial tidal field encountered by a region with scale $k$. This would determine the axis ratios governing the ellipsoidal drifting motion. Subsequently, the density evolution within this region would be subject to the contraction or expansion of each axis induced by the particle motion, described by
\begin{equation}
\frac{\rho}{\bar{\rho}_{\rm m}} = \prod_{i=1}^{3} \left|1 - \lambda_{i}\delta(k, a)\right|^{-1},
\end{equation}
where $\bar{\rho}_{\rm m}$ represents the average density of dark matter content. Also, $\lambda_{i}$'s denote the eigenvalues of the deformation tensor in the Zeldovich approximation, which can be obtained as
\begin{equation}
\lambda_{1}=\frac{1+3e+p}{3}, \hspace*{0.2cm}\lambda_{2}=\frac{1-2p}{3}, \hspace*{0.2cm}\lambda_{3}=\frac{1-3e+p}{3}.
\end{equation}
Along any given axis $i$, collapse occurs when the linear density perturbation $\delta(k,a)$ grows to exceed $\lambda_{i}^{-1}$. The axis with the smallest eigenvalue will be the last to collapse. Labeling this smallest eigenvalue by $\lambda_{3}$, the entire ellipsoidal region will collapse once $\lambda_{3}\delta(k, a_{\rm c})=1$, where $a_{\rm c}$ refers to the collapse scale factor. Under these conditions, the critical density contrast can be determined as
\begin{equation}\label{eq_delta}
\delta_{\rm c} = \frac{3}{1-3e+p}.
\end{equation}
In \cite{2001MNRAS.323....1S} it is demonstrated that in a Gaussian random field with a density contrast $\delta$ and a linear root-mean-square $\sigma$, the most probable values for eccentricity and prolateness would be $e_{\rm mp}=\sigma/(\sqrt{5}\delta)$ and $p_{\rm mp}=0$, respectively. These most probable values give the collapse threshold $\delta_{\rm c}=3(1+\sigma/\sqrt{5})$, which in the excursion set formalism translates into the moving barrier $B(S) = 3(1+\sqrt{S/5})$ with $S\equiv \sigma^{2}$ \citep{1991ApJ...379..440B}. Therefore, if the density is high enough, the collapse will occur during the radiation-dominated era. This is despite the fact that common mild perturbations can form abundant UDMHs, while only the rarest spikes are qualified to form PBHs.

One should note that halos are not guaranteed to form from collapsed structures in the radiation-dominated epoch. It has been shown by numerical simulations that a peak's collapse leads to the formation of a virialized halo only when the collapsed area becomes locally matter-dominated \citep{2019PhRvD.100j3010B}. On the other hand, the collapsed regions are found to be orders of magnitude denser than the dark matter background, with their densities scaling as $\sim e^{-2}\bar{\rho}_{\rm m}$ \citep{2023MNRAS.520.4370D}. Given the standard ellipticity $e\sim0.15$ of the tidal field at a $3\sigma$ peak, local matter dominance occurs at $a/a_{\rm eq} \sim e^{2} \simeq \mathcal{O}(10^{-2})$. This results in the formation of a halo well before the onset of matter domination. The density within these halos would be exceptionally high as they scale with the mean cosmic density at their formation time.
\subsection{Halo mass functions}\label{sec:mass-functions}
The excursion set theory provides the unconditional mass function of halos, which describes the average comoving number density of halos within a logarithmic mass interval \citep{1991ApJ...379..440B},
\begin{equation}
\frac{{\rm d}n}{{\rm d}\log M} = \frac{\bar{\rho}_{\rm m,0}}{M} \left|\frac{{\rm d}\log\nu}{{\rm d}\log M} \right| \nu f(\nu).
\end{equation}
Here, $\bar{\rho}_{\rm m,0} \simeq 33\, {\rm M_{\odot}/kpc^{3}}$ denotes the comoving average density of dark matter content, and $\nu \equiv \delta_{\rm c}/\sigma$ is a dimensionless parameter known as peak height. Also, the function $f(\nu)$, known as the ``multiplicity function'', represents the distribution of first crossings. As previously stated, $\sigma(M, a)$ represents the root-mean-square of linear density fluctuations smoothed on the mass scale $M$, i.e.,
\begin{equation}
\sigma^2(M, a) = \frac{1}{2\pi^{2}}\int_{0}^{\infty} P(k, a) W^2(k, M) k^{2} {\rm d}k,
\end{equation}
where $P(k, a)$ refers to the power spectrum of linear matter fluctuations and $W(k, M)$ is the smoothing window function, assumed to be a sharp$\mbox{-}k$ filter\footnote{For the small-scale power spectrum we are interested in, employing a real-space top-hat window function with a truncated power spectrum may result in specific issues, such as increased variance even for masses below the cutoff scale. This can be addressed by utilizing the sharp-$k$ window function \citep{2020JCAP...12..038T, 2022ApJ...928L..20S, 2023A&A...669L...2A}.} \citep{1991ApJ...379..440B},
\begin{equation}
W(k, M) = 
\left\{
\begin{array}{ll}
1 & \hspace*{0.5cm}\text{if } \,\, 0 < k \leq k_{M}\\
0 & \hspace*{0.5cm}\rm{otherwise}
\end{array}
\right.
\end{equation}
with $k_{M}=(6\pi^{2}\bar{\rho}_{\rm m,0}/M)^{1/3}$ \citep{1994MNRAS.271..676L}. 
Under such conditions we have
\begin{equation}
\frac{{\rm d}\log \nu}{{\rm d}\log M}=\frac{k_{M}^{3}}{12\pi^{2}}\frac{P(k_{M}, a)}{\sigma^{2}(M, a)}.
\end{equation}
From Eq.\,(\ref{eq1}) one gets
\begin{equation}
P(k, a)=\frac{2\pi^{2}\alpha^{2}}{k^{3}}\left[\log\left(\sqrt{2}\beta \frac{k}{k_{\rm eq}}\frac{a}{a_{\rm eq}}\right)\right]^{2}P_{\zeta}(k),
\end{equation}
where $P_{\zeta}(k)$ is the dimensionless power spectrum of primordial curvature perturbations. The exact form of $P_{\zeta}(k)$ depends on the assumed inflationary scenario that seeds the perturbations. We will provide some suggestions in Section~\ref{sec:powerspectra}.

Our main goal in this work is to explore the role of certain physical factors in the halo mass function in the ultradense regime within the excursion set theory, through their impact on the dynamical barrier. In scenarios with different types of barriers, the distribution of first crossings can be determined by simulating a significant number of random walks \citep{1998MNRAS.300.1057S, 2002MNRAS.329...61S}. In \cite{2002MNRAS.329...61S} it is shown that, for a wide range of dynamical barriers, the initial crossing distribution can be approximated by
\begin{equation} 
f(S) = |T(S)|\exp\left(-\frac{B(S)^{2}}{2S}\right)\frac{1}{S\sqrt{2\pi S}},
\end{equation}
where $T(S)$ is given by the Taylor expansion of $B(S)$
\begin{equation}
T(S) = \sum_{n=0}^{5}\frac{(-S)^{n}}{n!}\frac{\partial^{n}B}{\partial S^{n}}.
\end{equation}
\cite{2023MNRAS.520.4370D} derive an approximation for the distribution of first barrier crossings within a Gaussian random walk framework, validated by Monte Carlo simulation,
\begin{equation}
f(S) \approx \frac{3 + 0.556\sqrt{S}}{\sqrt{2\pi S^{3}}} \exp\left(-\frac{B^2}{2S}\right) \left(1 + \frac{S}{400}\right)^{-0.15},
\end{equation}
which corresponds to the PS multiplicity function\footnote{The statement explains how the multiplicity function $f(\nu)$ is connected to the first crossing distribution $f(S,t)$. The latter represents the probability distribution for the first time a random walk crosses a specific barrier $B(S)$. The equation $\nu f(\nu) = Sf(S,t)$ demonstrates that $f(\nu)$ can be derived from $f(S,t)$, illustrating a fundamental relation within the excursion set theory \citep{2007IJMPD..16..763Z}.},
\begin{equation}
\left[\nu f(\nu)\right]_{\rm PS} = \sqrt{\frac{2}{\pi}}\frac{(\nu + 0.556) \exp\left[-0.5(1+\nu^{1.34})^2\right]}{(1+0.0225\nu^{-2})^{0.15}}.
\end{equation}

There are, however, discrepancies between the predictions of the PS mass function and dark matter halo distributions, especially at high redshifts \citep{2017JCAP...03..032D}. The disagreements, as already stated, might be attributed to various physical factors ignored in the PS formalism and may have a non-negligible impact on the abundance of UDMHs. The first factor considered here is geometrical and generalizes the spherical-collapse model in the essence of PS formalism to ellipsoidal-collapse halo models. The modified mass function is known as Sheth-Tormen (ST), described by \citep{2001MNRAS.323....1S}
\begin{equation}
\left[\nu f(\nu) \right]_{\rm ST} = A_{1}\sqrt{\frac{2 \nu^{\prime}}{\pi}}\left(1+\frac{1}{\nu^{\prime q}}\right)\exp\left(-\frac{\nu^{\prime}}{2}\right),
\end{equation}
where $q=0.3$, $\nu^{\prime} \equiv 0.707 \nu^{2}$, and $A_{1}=0.322$ is determined by demanding that the integral of $f(\nu)$ across all possible values of $\nu$ equals unity.

Besides the geometric conditions during the virialization process of dark matter halos, other physical factors can also impact the collapse of overdense regions and, consequently, the halo mass function. Incorporating these essential physical factors into the analysis is crucial as they capture the genuine physics governing halo collapse and growth and the processes underlying structure formation and evolution throughout cosmic history. Additionally, this approach allows the collapse threshold to depend on effective physical factors. As a result, the barrier adjusts accordingly with these variables, resulting in a more realistic model for halo collapse. Among these corrections, we consider the impact of angular momentum, dynamical friction, and cosmological constant on the halo mass function. These corrections are shown to play a role in reducing the discrepancies, particularly in controversial mass ranges. Including the effect of angular momentum and cosmological constant \footnote{It should be stressed that the cosmological constant is not expected to play any role in the high redshifts we are interested in. However, we retain it for completeness.}, the mass function, referred to as DP1 in this work, is found to be \citep{2006ApJ...637...12D}
\begin{equation}
\left[\nu f(\nu)\right]_{\rm DP1} \approx A_{2} \sqrt{\frac{\nu^{\prime}}{2\pi}} k(\nu^{\prime}) \exp\{-0.4019 \nu^{\prime}l(\nu^{\prime})\},
\end{equation}
with $A_{2} = 0.974$ set by normalization, and
\begin{equation}
k(\nu^{\prime}) = \left(1 + \frac{0.1218}{(\nu^{\prime})^{0.585}} + \frac{0.0079}{(\nu^{\prime})^{0.4}} \right),
\end{equation}
and
\begin{equation}
l(\nu^{\prime}) = \left(1 + \frac{0.5526}{(\nu^{\prime})^{0.585}} + \frac{0.02}{(\nu^{\prime})^{0.4}} \right)^{2}.
\end{equation}

The influence of dynamical friction on the barrier was also investigated in \cite{2017JCAP...03..032D}, leading to the (hereafter DP2) mass function, 
\begin{equation}
\left[\nu f(\nu)\right]_{\rm DP2} \approx A_{3} \sqrt{\frac{\nu^{\prime}}{2\pi}}m(\nu^{\prime}) \exp\{-0.305 \nu^{\prime 2.12}n(\nu^{\prime})\},
\end{equation}
where $A_{3} = 0.937$ is set by normalization, and we have
\begin{equation}
m(\nu^{\prime}) = \left(1 + \frac{0.1218}{(\nu^{\prime})^{0.585}} + \frac{0.0079}{(\nu^{\prime})^{0.4}} + \frac{0.1}{(\nu^{\prime})^{0.45}}\right),
\end{equation}
and
\begin{equation}
n(\nu^{\prime}) = \left(1 + \frac{0.5526}{(\nu^{\prime})^{0.585}} + \frac{0.02}{(\nu^{\prime})^{0.4}} + \frac{0.07}{(\nu^{\prime})^{0.45}}\right)^{2}.
\end{equation}
We will use these mass functions in Section~\ref{sec:iii} to calculate the abundance of UDMHs for various primordial power spectra introduced in the next section. 
\subsection{Primordial power spectra}\label{sec:powerspectra}
Formation of PBHs and UDMHs in the radiation-dominated era requires substantial amplification (of the order of $\mathcal{O}(10^{7})$) of the primordial power spectrum on small scales compared to the observationally-supported almost scale-invariant spectrum on large scales. We therefore consider modifications to this scale-invariant spectrum, $P_{\zeta_{0}}(k)$, by some large-amplitude fluctuations on small scales, $P_{\zeta_{1}}$, i.e., 
\begin{eqnarray} \label{powerspec1}
P_\zeta^{\rm x}(k) = P_{\zeta_{0}}(k) + P_{\zeta_{1}}^{\rm x}(k),
\end{eqnarray}
where $P_\zeta^{\rm x}(k)$ is the total spectrum, ``x" stands for the different models used, and
\begin{eqnarray}
P_{\zeta_{0}} = A_{0}\left(\frac{k}{k_{\rm{p}_{0}}}\right)^{n_{\rm s}-1}.
\end{eqnarray}
Here $k_{\rm{p}_{0}}=5\times 10^{-2}\,\mathrm{Mpc}^{-1}$ is the scalar pivot scale and $A_{0} = 2.1 \times 10^{-9}$ and $n_{\rm s} = 0.96$ are set from cosmic microwave background observations \citep{2020A&A...641A...6P}. In this work, we explore four models for $P_{\zeta_{1}}^{\rm x}(k)$ with different features to allow for PBH and UDMH formation on various mass scales. 

As the first extension, we consider a smooth transition from large- to small-scale spectrum, also assumed 
to be nearly scale-invariant on these scales, 
\begin{eqnarray}\label{pzetas}
P_{\zeta_{1}}^{\rm S} = A_{1}\left(\frac{k}{k_{{\rm p}_{1}}}\right)^{n_{\rm s}-1}D(k, k_{1}, w) \exp\left[\frac{-(k-k_{2})^{2}}{2\sigma^{2}}\right],\nonumber\\
\end{eqnarray}
where $k_{\rm{p}_{1}}=10^6\, \mathrm{Mpc}^{-1}$ and $A_{1} \approx 2.2 \times 10^{-2}$ are chosen as the pivot scale and amplitude for the small-scale power spectrum, and we have defined 
\begin{eqnarray}
D(k, k_{1}, w) = \frac{1}{2}\left[1 +\tanh\left(\frac{k - k_{1}}{w}\right)\right]
\end{eqnarray}
as the smoothing function with the transition scale $k_1 \simeq 190$ Mpc$^{-1}$ and width $w=15$ Mpc$^{-1}$. The Gaussian function is used to smoothly kill the amplified fluctuations on very small scales beyond the cut-off wavenumber $k_{2} \simeq 6\times 10^{4}$ Mpc$^{-1}$ with a width $\sigma=2\times 10^{5}$ Mpc$^{-1}$. The maximum mass of UDMHs seeded by this power spectrum would then be automatically set to $\sim 10^{5}\,M_{\odot}$.

\begin{figure}
\centering
\includegraphics[width=0.9\linewidth]{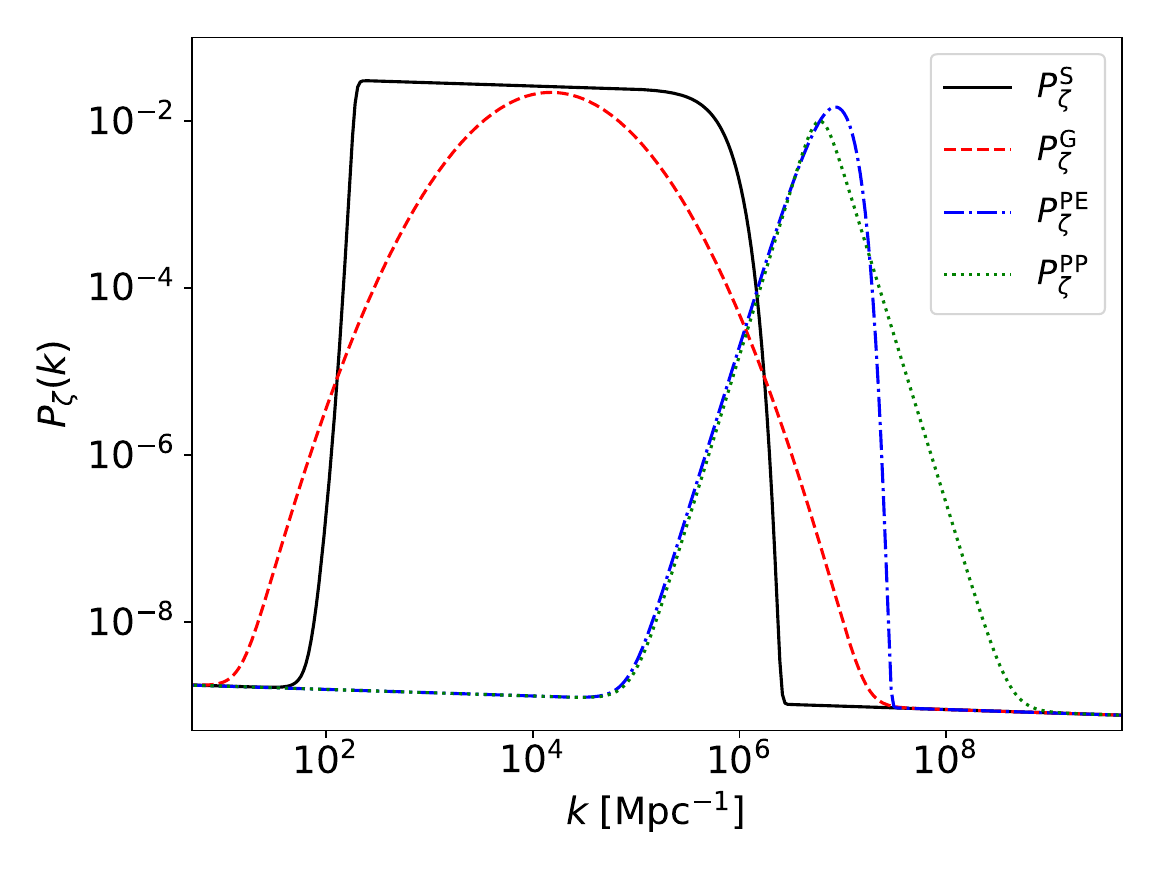}
\caption{Primordial power spectra as a function of $k$ for the four models introduced in Section~\ref{sec:powerspectra}. The (black) sloid, (red) dashed, (blue) dot-dashed, and (green) dotted lines correspond to the spectra described by Eq.~(\ref{powerspec1}), with the amplifications given by Eqs.\,(\ref{pzetas}), (\ref{pzetag}), (\ref{ppezeta}), and (\ref{pppzeta}), respectively. The characteristic scales are chosen to be $k_1=190\,{\rm Mpc^{-1}}$ for $P_{\zeta}^{\rm S}$, $1.5\times 10^{4}\,\rm{Mpc}^{-1}$ for $P_{\zeta}^{\rm G}$, and $6\times 10^{6}\,{\rm Mpc^{-1}}$ for $P_{\zeta}^{\rm PE}$ and $P_{\zeta}^{\rm PP}$. The small-scale amplitude of all power spectra is set to be $A_1\approx 2.2\times 10^{-2}$.}
\label{fig_1}
\end{figure}

\begin{figure*}[htbp]
\includegraphics[width=0.33\linewidth]{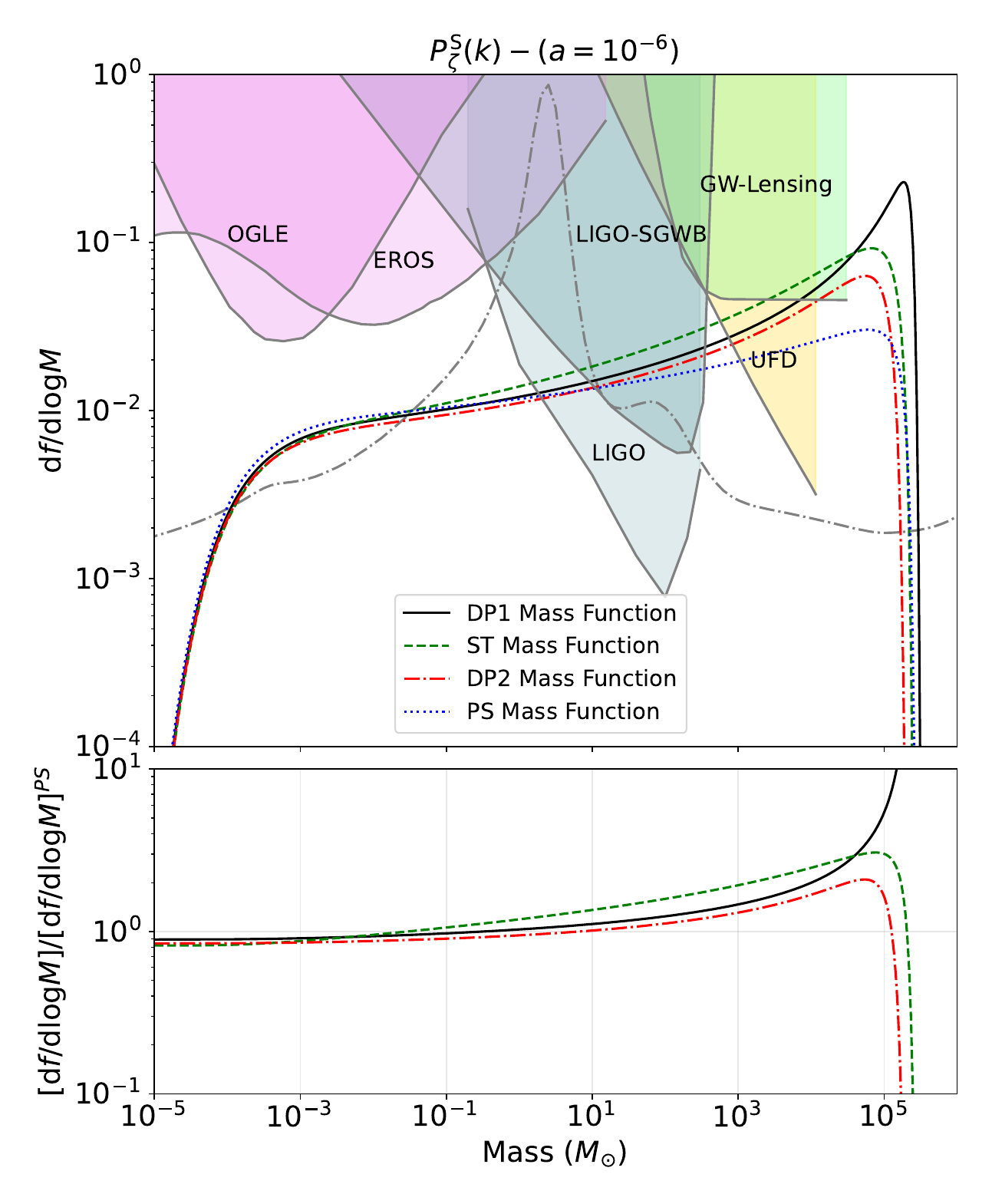}
\includegraphics[width=0.33\linewidth]{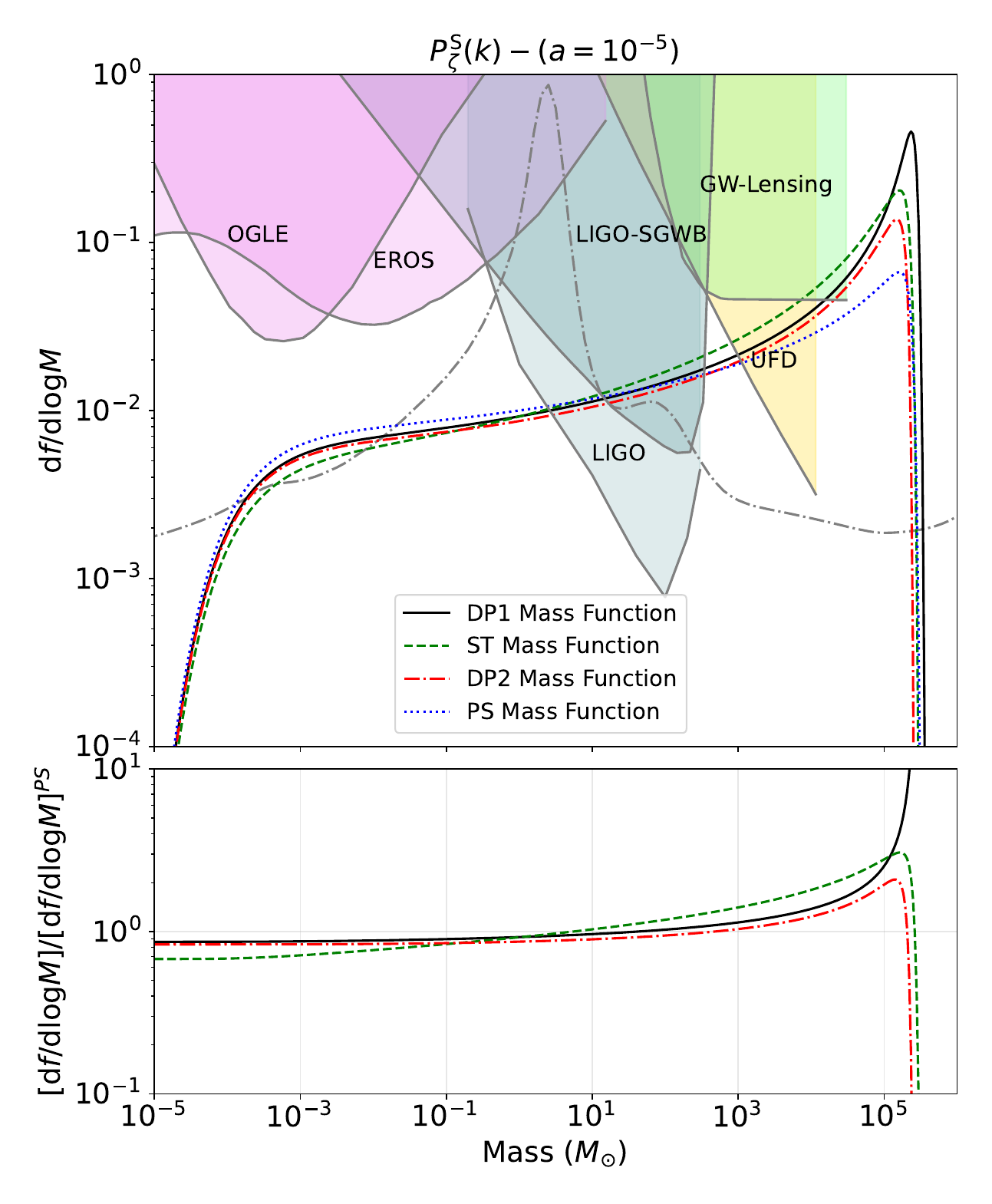}
\includegraphics[width=0.33\linewidth]{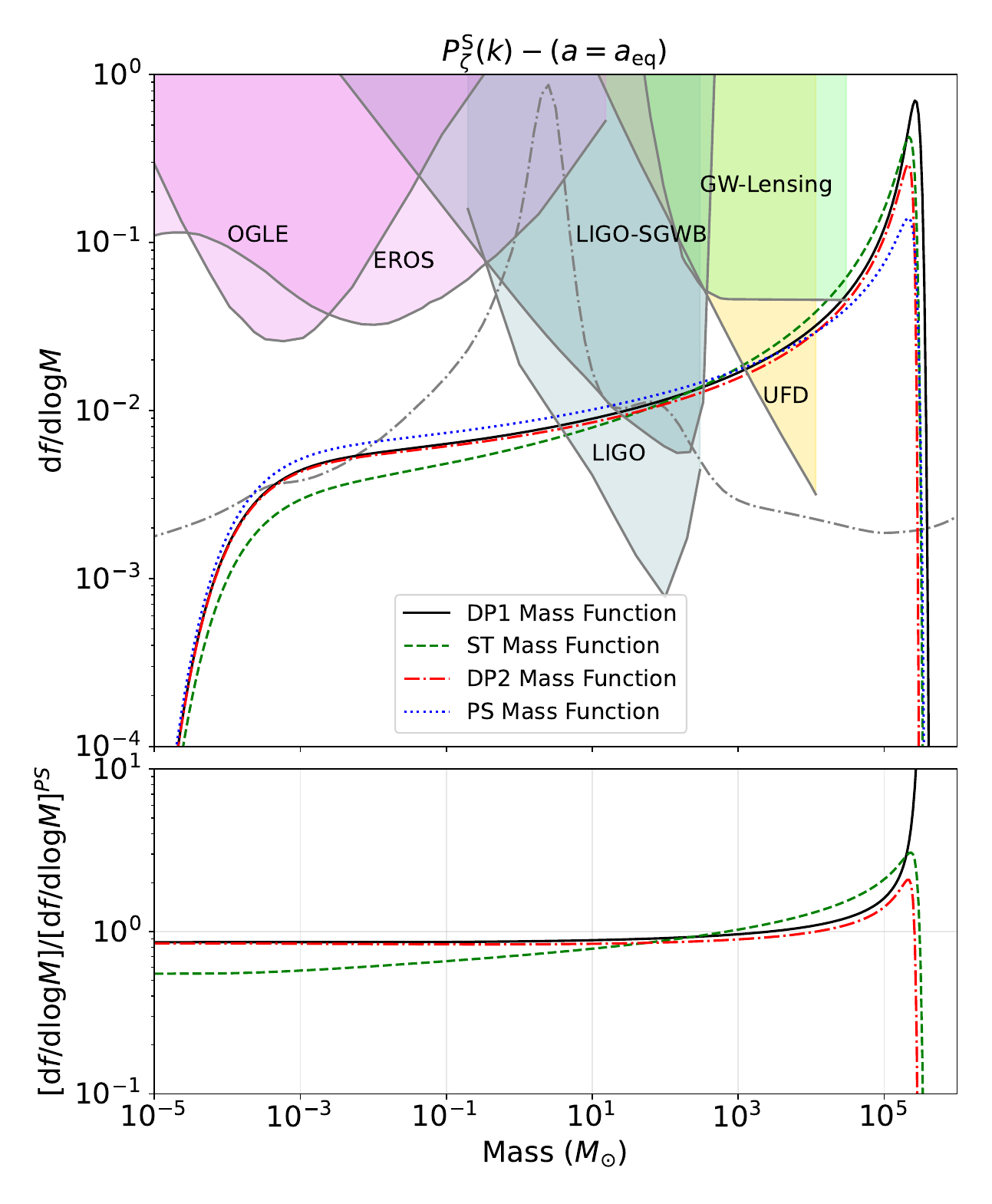}
\caption{Top: differential dark matter mass fraction in UDMHs for different scale factors $(a\leq a_{\rm eq})$, seeded by primordial perturbations described by the power spectrum $P_{\zeta}^{\rm S}(k)$. The (black) solid, (green) dashed, (red) dot-dashed, and (blue) dotted lines represent the results for DP1, ST, DP2, and PS mass functions. The (gray) dot-dashed line indicates the mass spectrum of PBHs derived from the thermal history of the early Universe, calculated for the smoothed power spectrum of Eq.\,(\ref{pzetas}) and for $f_{\rm PBH}=1$. Observational constraints on PBHs are plotted to compare the predicted abundance of UDMHs and the upper limits on PBHs while considering the extended mass spectrum. The constraints are based on the results from microlensing data by EROS \citep{2007A&A...469..387T}, the disruption of ultra-faint dwarfs (UFD) \citep{2016ApJ...824L..31B}, microlensing of GWs by PBHs (GW-Lensing) \citep{2019PhRvL.122d1103J}, microlensing data by OGLE \citep{2019PhRvD..99h3503N}, stochastic background of PBH mergers with LIGO (LIGO-SGWB) \citep{2020JCAP...08..039C}, and direct constraints on PBH-PBH mergers with LIGO \citep{2019PhRvD.100b4017A, 2022PhRvL.129f1104A}. Bottom: relative deviation of ST, DP1, and DP2 mass functions compared to the PS mass function.}
\label{fig_2}
\end{figure*}
As the second extension, we model a localized amplification of the power spectrum in the $k$-space with a Gaussian bump,
\begin{eqnarray}\label{pzetag}
P_{\zeta_{1}}^{\rm{G}} = A_{1}\exp\left(- \frac{\log^{2}(k/k_{1})}{2\sigma^{2}}\right),
\end{eqnarray}
where the bump is centered at $k_{1}=1.5\times 10^{4}\, \mathrm{Mpc}^{-1}$ with a width of $\sigma=1.2$ for the extended distribution, and $\sigma=5\times 10^{-2}$ for the narrow distribution. Additionally, the amplitude is set to $A_{1}=10^{-2}$.

As the third model we consider a certain combination of power-law and exponential function as \citep{2023PhRvD.107h3505D} 
\begin{eqnarray}\label{ppezeta}
P_{\zeta_{1}}^{\rm PE}(k) = A_{1}\left(\frac{k}{k_{1}}\right)^{4}\exp\left[1-\left(\frac{k}{k_{1}}\right)^{2}\right],
\end{eqnarray}
where $A_{1}=P_{\zeta_{1}}^{\rm PE}(k_{1})$ is the amplitude of the spectrum at the characteristic scale $k_{1}$. The spectrum experiences an expansion proportional to $k^{4}$ for $k<k_{1}$, representing the distinctive steep growth observed in conventional ultraslow-roll inflation scenarios \citep{2019JCAP...06..028B, 2022PhRvD.106l3519F, 2023JCAP...03..013K}. For $k>k_{1}$ the spectrum undergoes damping.

The last case we consider is constructed from a combination of two power-laws as
\begin{eqnarray}\label{pppzeta}
P_{\zeta_{1}}^{\rm PP}(k) = 2A_{1}\left[\left(\frac{k}{k_{1}}\right)^{-4} + \left(\frac{k}{k_{1}}\right)^{4}\right]^{-1},
\end{eqnarray}
with a damping tail different from $P_{\zeta}^{\rm PE}$. 
Figure\,\ref{fig_1} illustrates the four power spectra introduced in this section. As can be seen, the amplitudes are chosen so that the spectra are amplified to $\sim 10^{-2}$ at the characteristic scale $k_{1}$ of each model. 

\section{Results and discussions}\label{sec:iii}
\begin{figure*}[htbp]
\includegraphics[width=0.33\linewidth]{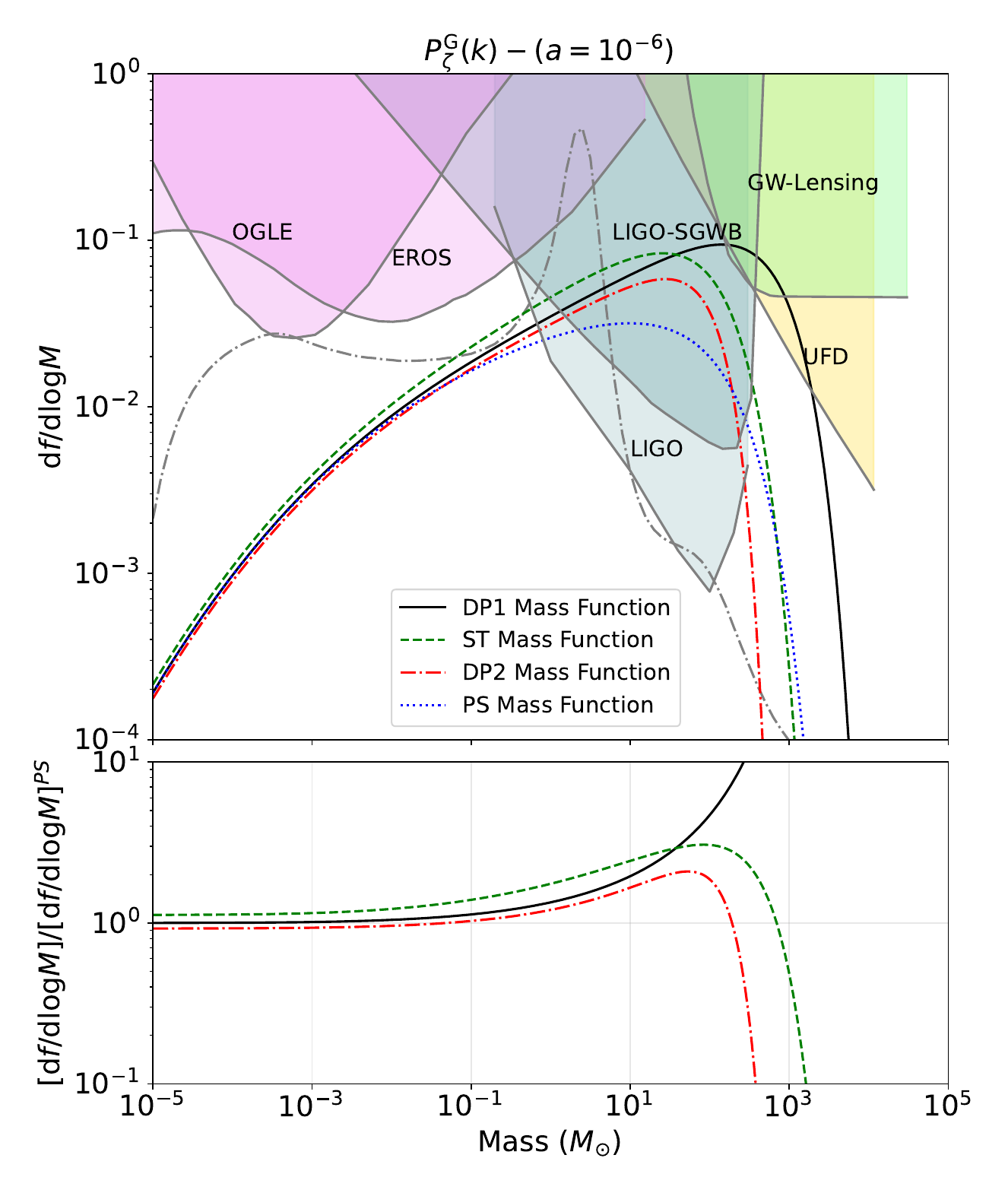}
\includegraphics[width=0.33\linewidth]{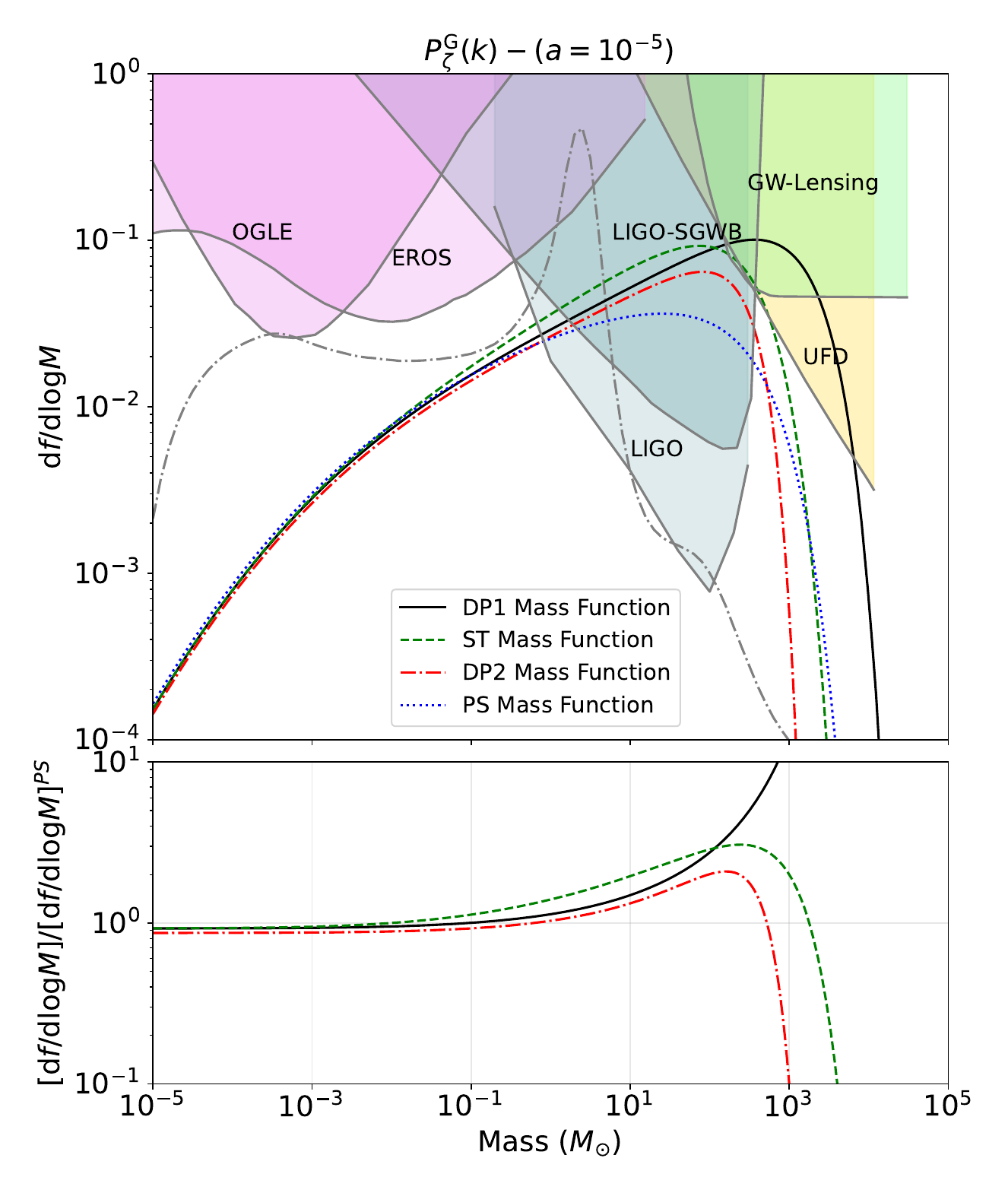}
\includegraphics[width=0.33\linewidth]{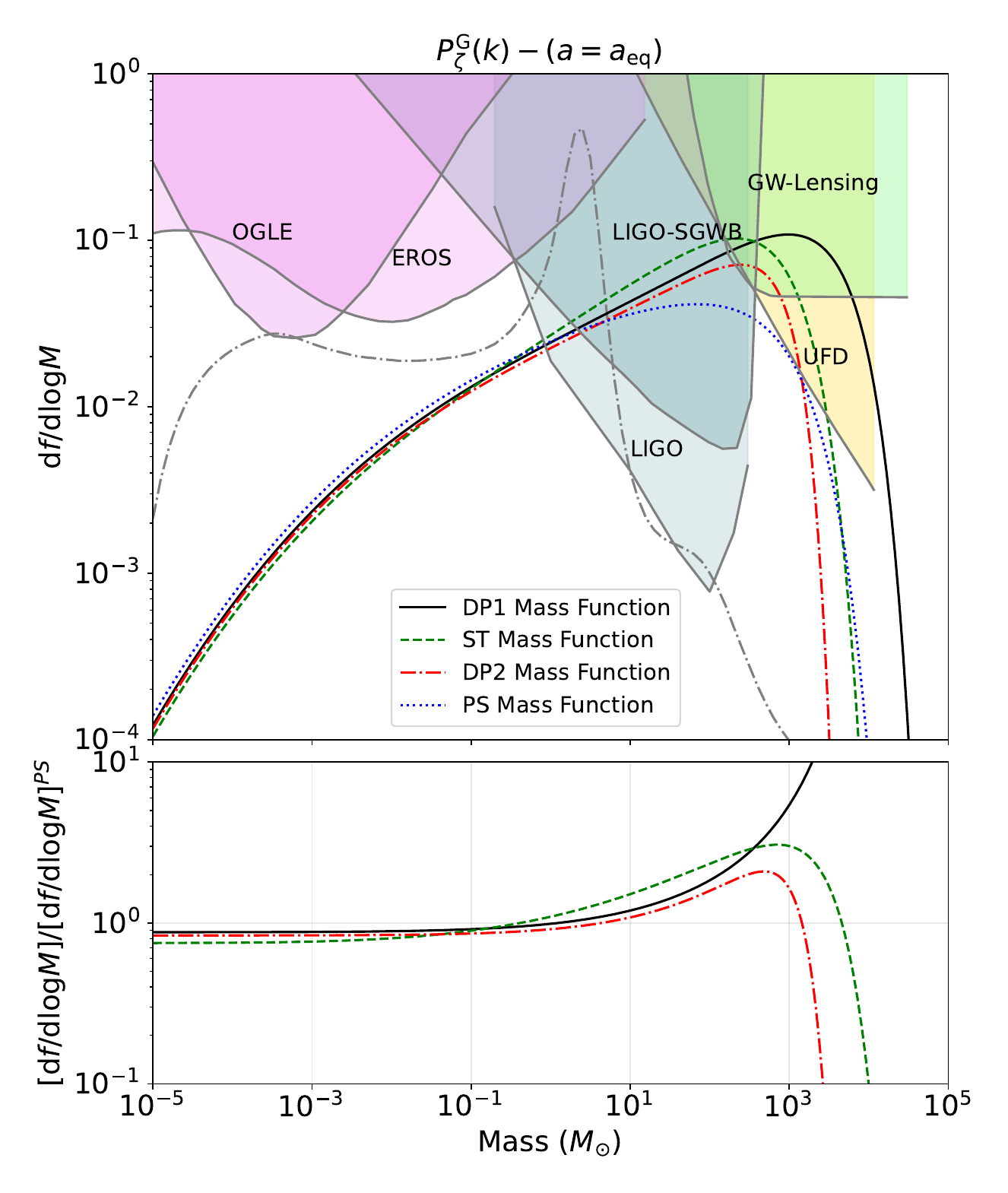}
\caption{Similar to Figure\,\ref{fig_2} but for the extended Gaussian primordial power spectrum. The (gray) dot-dashed line indicates the mass spectrum of PBHs derived from the thermal history of the early Universe, while considering Eq.\,(\ref{pzetag}) with $\sigma=1.2$.} 
\label{fig_3}
\end{figure*}

\begin{figure*}[htbp]
\includegraphics[width=0.33\linewidth]{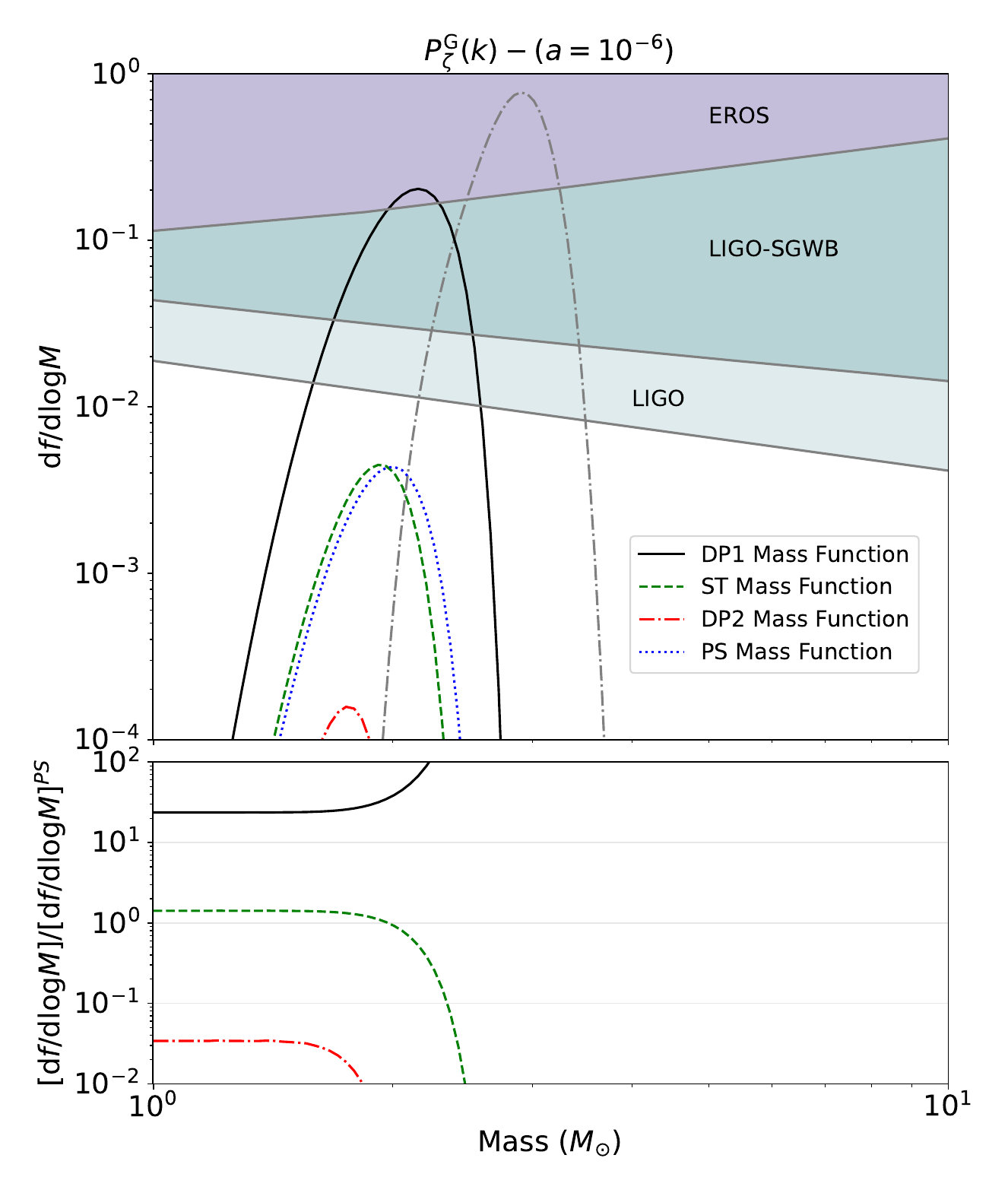}
\includegraphics[width=0.33\linewidth]{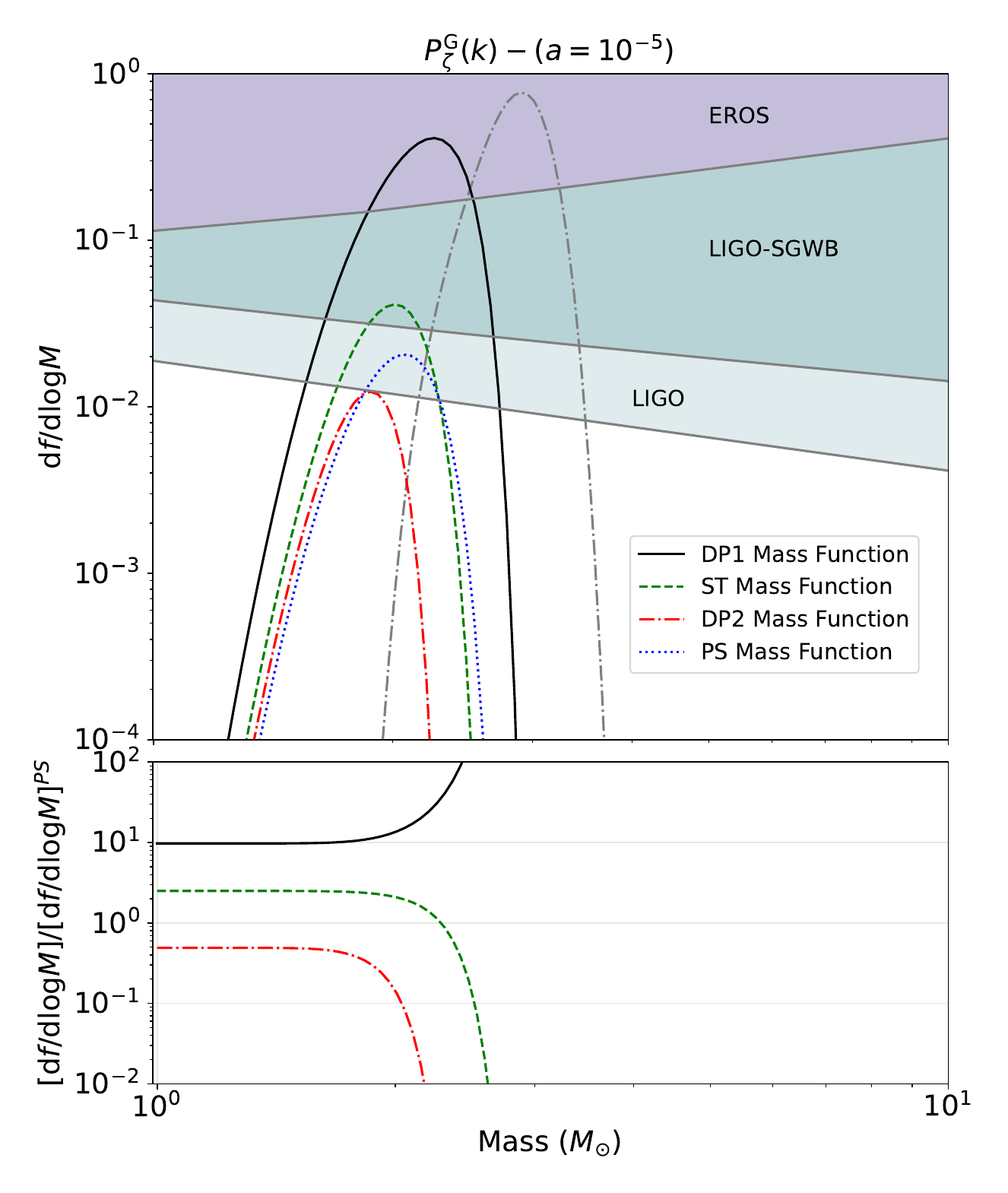}
\includegraphics[width=0.33\linewidth]{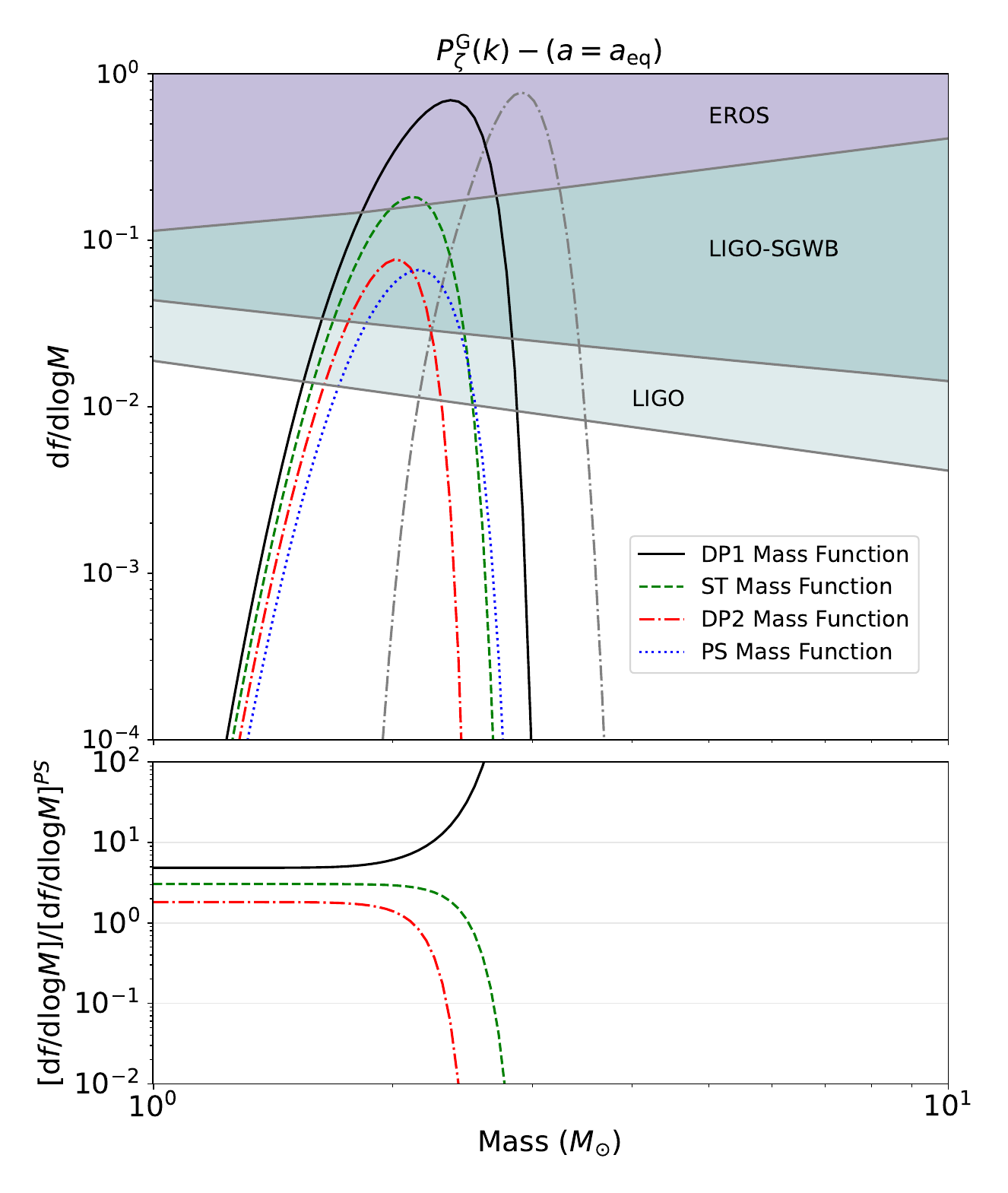}
\caption{Similar to Figure\,\ref{fig_2} but for the narrow Gaussian power spectrum. The (gray) dot-dashed line indicates the mass spectrum of PBHs derived from the thermal history of the early Universe, while considering Eq.\,(\ref{pzetag}) with $\sigma=5\times 10^{-2}$.} 
\label{fig_4}
\end{figure*}

\begin{figure*}[htbp]
\centering
\includegraphics[width=0.33\linewidth]{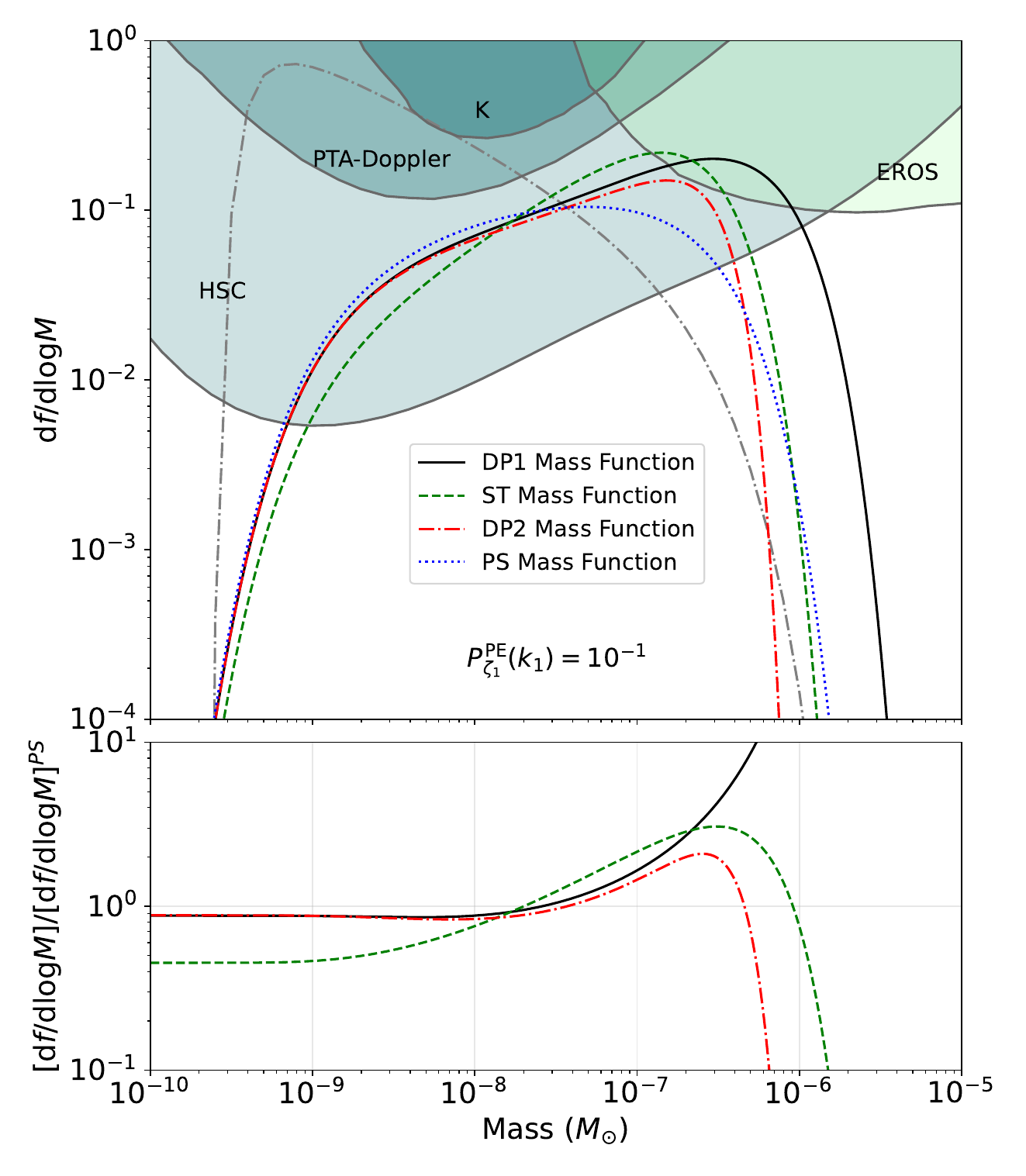}
\includegraphics[width=0.33\linewidth]{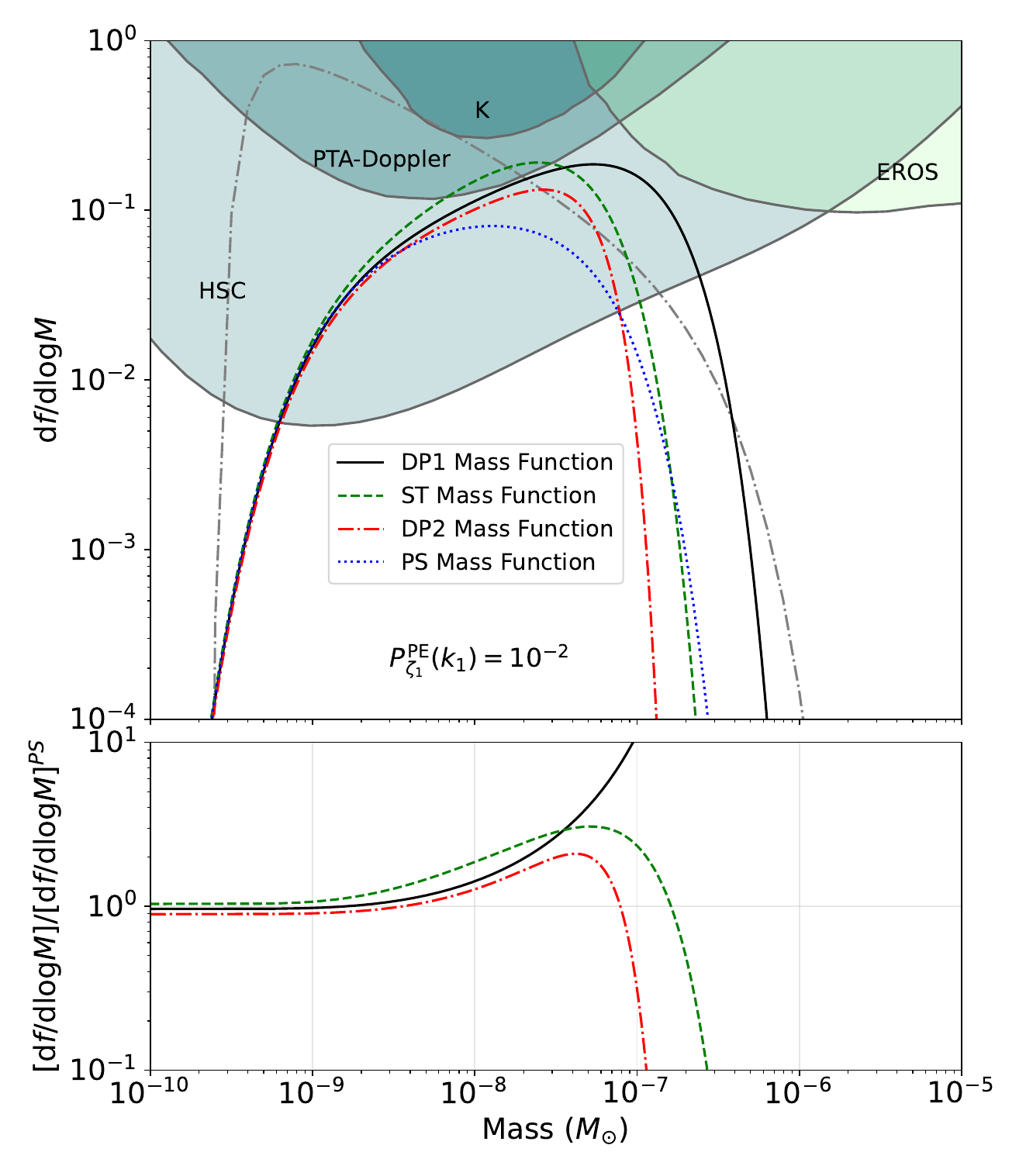}
\includegraphics[width=0.33\linewidth]{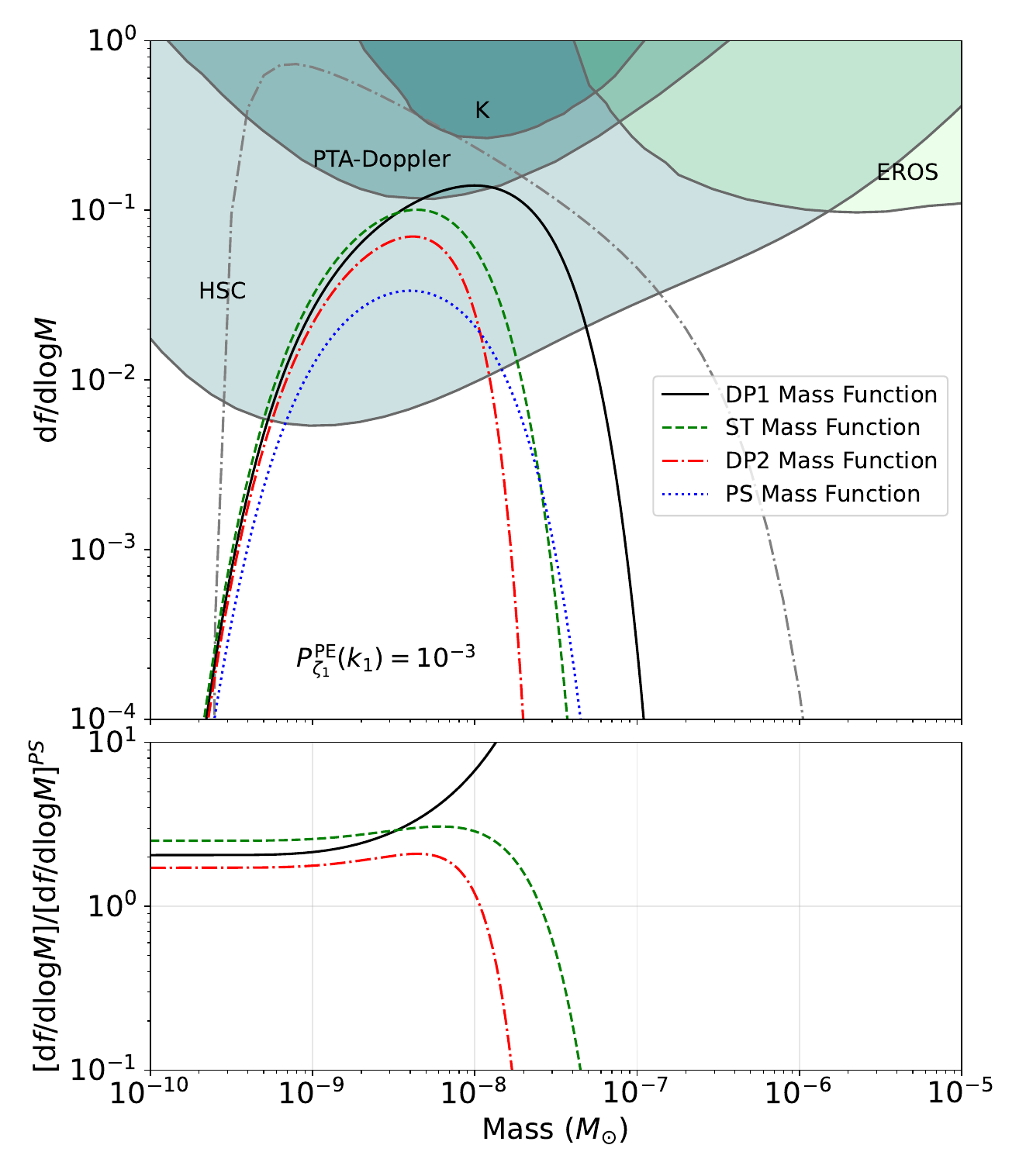}
\caption{Differential mass fraction in UDMHs as a function of mass $M$, with $P_{\zeta}^{\rm PE}(k)$ as the primordial power spectrum at $a_{\rm eq}$, and for different primordial amplitudes (from left to right: $P_{\zeta_{1}}^{\rm PE}(k_{1})=10^{-1}$,$10^{-2}$ and $10^{-3}$, respectively). The (black) solid, (green) dashed, (red) dot-dashed, and (blue) dotted lines correspond to the results for DP1, ST, DP2, and PS mass functions. The characteristic scale $k_{1}$ is set to $6\times 10^{6}\,{\rm Mpc^{-1}}$. The (gray) dot-dashed line indicates the mass spectrum of PBHs derived from the thermal history of the early Universe, with the power spectrum defined in Eq.\,(\ref{ppezeta}), and assuming $f_{\rm PBH}=1$. For comparison with the bounds on the PBH abundance, observational constraints are also plotted, including microlensing constraints from EROS \citep{2007A&A...469..387T}, microlensing constraints from Kepler (K) \citep{2014ApJ...786..158G}, PBH-induced motions in pulsar timing arrays (PTA-Doppler) \citep{2019PhRvD.100b3003D}, and microlensing constraints from HSC \citep{2020PhRvD.102h3021C}.}
\label{fig_5}
\end{figure*}

\begin{figure*}[htbp]
\centering
\includegraphics[width=0.33\linewidth]{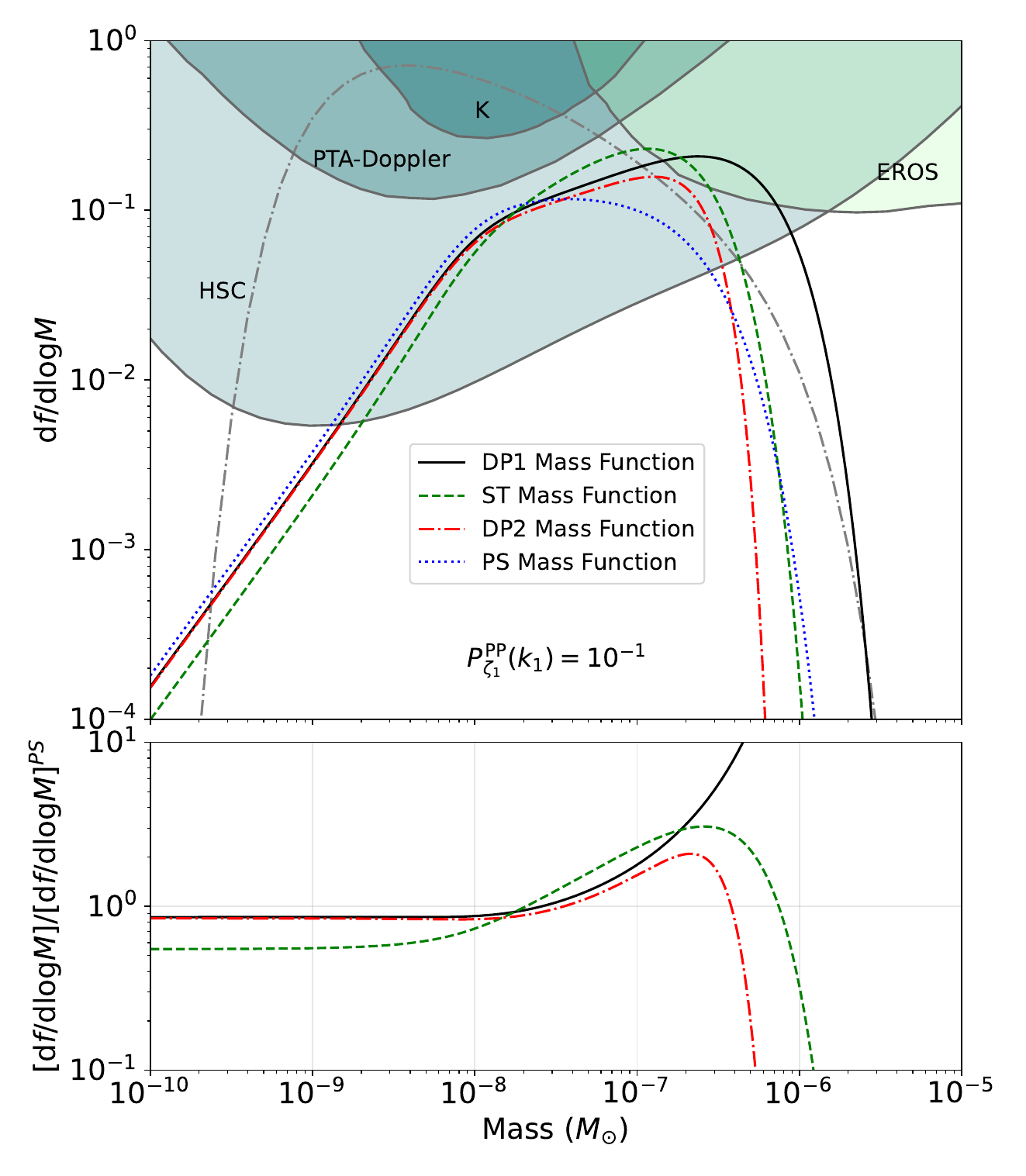}
\includegraphics[width=0.33\linewidth]{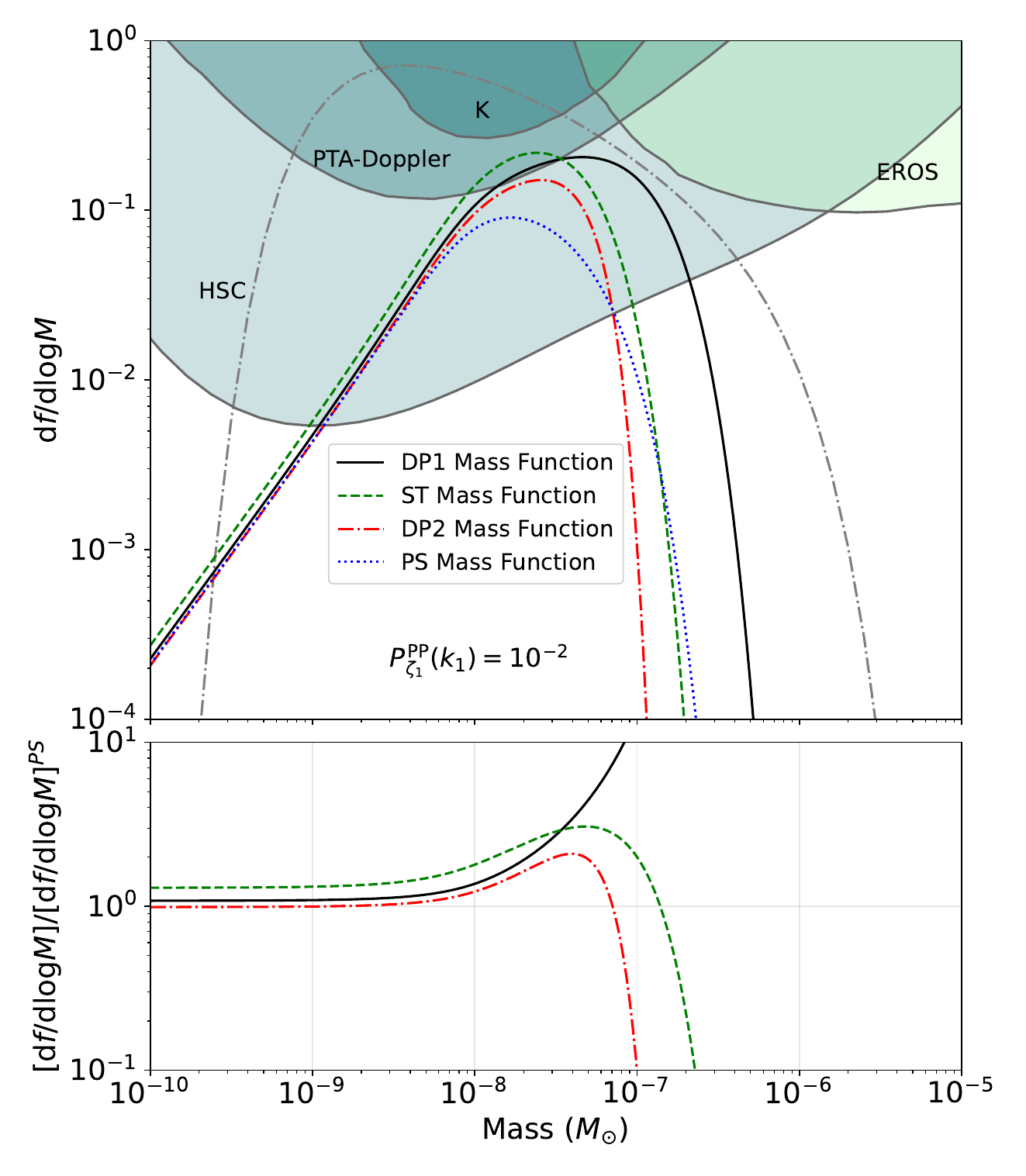}
\includegraphics[width=0.33\linewidth]{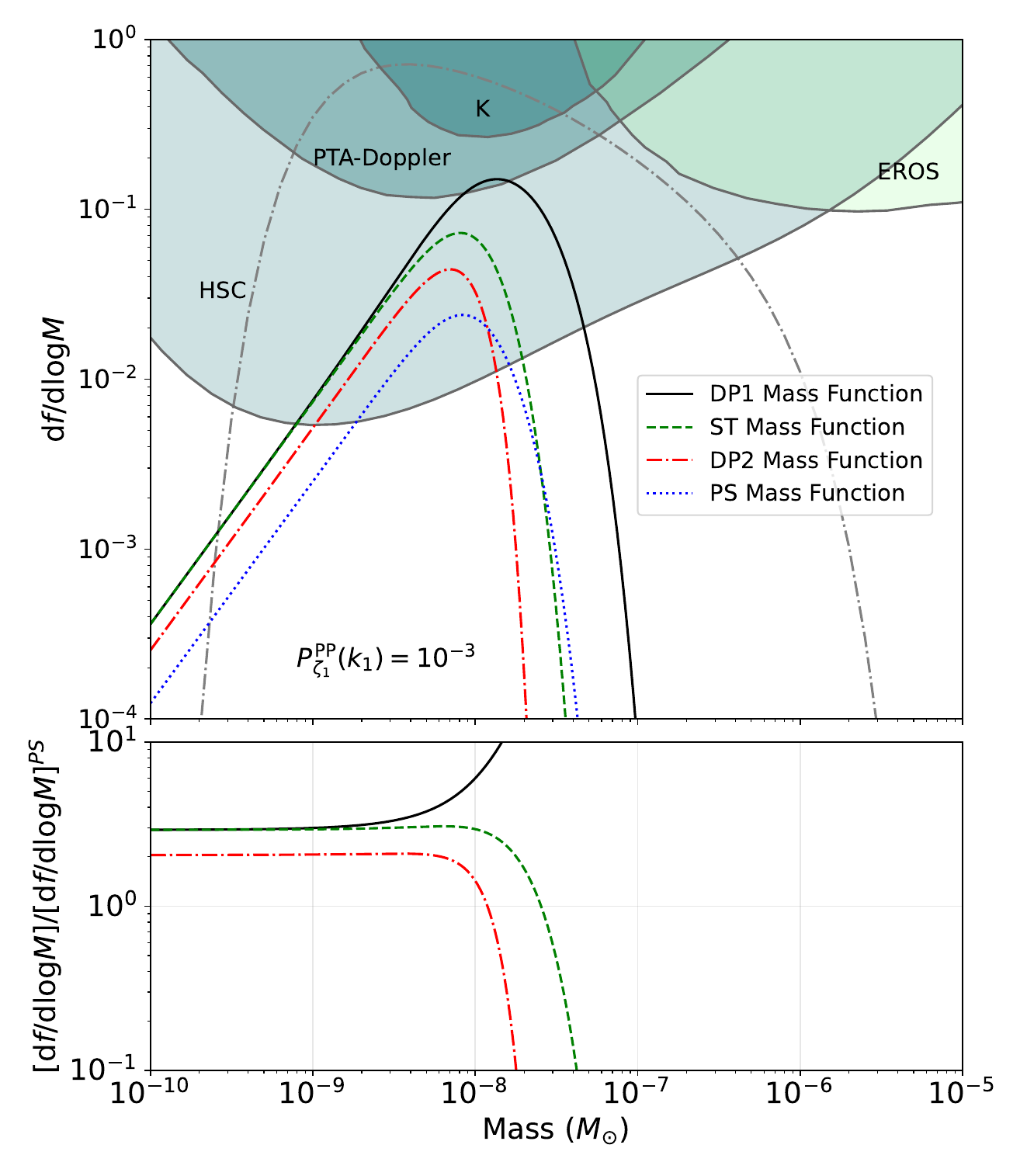}
\caption{Similar to Figure\,\ref{fig_5} but with $P_{\zeta}^{\rm PP}(k)$ as the primordial power spectrum. The (gray) dot-dashed line indicates the mass spectrum of PBHs derived from the thermal history of the early Universe, with the primordial power spectrum modeled by Eq.\,(\ref{pppzeta}).}
\label{fig_6}
\end{figure*}

In this section we present and compare the estimated abundance of UDMHs based on the modifications to halo formation processes discussed in Section~\ref{sec:mass-functions}. Figures\,\ref{fig_2}--\ref{fig_6} illustrate the differential mass fraction ${\rm d}f/{\rm d}\log M = (M/\bar{\rho}_{\rm m,0}) ({\rm d}n/{\rm d}\log M)$ of UDMHs as a function of mass $M$ in the radiation-dominated epoch, i.e., $a\leq a_{\rm eq}$, with the various models of primordial power spectra of Section\,\ref{sec:powerspectra}. We have considered three distinct halo mass functions: the ST approximation, and DP1 and DP2 formulations, which are compared to the results from the PS formalism. Moreover, for a more accurate comparison, we have plotted three utilized mass functions normalized to the PS mass function as a function of mass $M$.

Figure \ref{fig_2} corresponds to the power spectrum with the smooth transition described by Eq.\,(\ref{powerspec1}). Recall that for the formation of UDMHs in the radiation-dominated era, ultradense regions must be locally matter-dominated. However, this condition is not expected to be satisfied earlier than $a/a_{\rm eq} \simeq \mathcal{O}( 10^{-2})$. Therefore, we have shown the results for $a=10^{-6}, 10^{-5}$, and $a_{\rm eq}$. As can be seen, the DP1, ST, and DP2 mass functions yield a higher abundance of large-mass UDMHs ($M \geq 10^{4}M_{\odot}$) when compared to the PS mass function, with DP1 predicting the highest abundance for these halos. 
This illustrates the impact of corrections due to angular momentum as the main physical factor considered in the DP1 scenario, since the cosmological constant is not expected to play a significant role in the early Universe. Including dynamical friction, on the other hand, reduces the formation of these dense halos, as is visible from the DP2 curve. 
The ST mass function, with its incorporation of triaxial collapse in the halo formation scenario, seems to maintain a rigorous standard and falls in between the previous two. 

To compare the relative abundance of UDMHs and accompanying PBHs, we examine two different situations. The SC scenario assumes that all dark matter consists entirely of PBHs, where $f_{\rm PBH}=1$. The MC scenario considers dark matter composed of distinct components, including particles and PBHs, with $f_{P\rm BH}<1$. In the SC scenario, we have provided the mass spectrum of PBHs based on assumptions for the thermal history of the early Universe \citep{2019arXiv190608217C}, derived for the specific power spectrum mentioned in Eq.\,(\ref{pzetas}). In this situation, the majority of dark matter is composed of $\mathcal{O}(1) M_{\odot}$ PBHs, while a smaller number of significantly larger PBHs serve as the initial seeds for the supermassive black holes located at the centers of galaxies. This comparison reveals that for masses exceeding $10 M_{\odot}$, a significant proportion of dark matter could be found in UDMHs, which originate from the DP1, ST, DP2, and PS mass functions, respectively. In this scenario, the formation of UDMHs likely results from the clustering of smaller PBHs. On the other hand, in the MC scenario, it is expected that the UDMHs originated from the virialization of dark matter particles, while the PBHs were formed through the direct collapse of density fluctuations. Hence, for comparison with the UDMHs abundance, we have plotted observational constraints on the abundance of PBHs for the extended mass functions \citep{2017PhRvD..96b3514C} \footnote{Note that the relevance of this comparison is limited to the specific circumstances outlined in the MC scenario. Moreover, an intriguing line of inquiry would be to compare predictions derived from more realistic mass functions with the observational constraints on UDMHs themselves, which is beyond the scope of this work.}. These constraints include microlensing constraints from EROS and OGLE \citep{2007A&A...469..387T, 2019PhRvD..99h3503N}, scalar-induced GWs with NANOGrav \citep{2020PhRvL.124y1101C}, direct constraints on PBH mergers with LIGO detector \citep{2019PhRvD.100b4017A, 2022PhRvL.129f1104A}, the stochastic background of PBH mergers with LIGO detector \citep{2020JCAP...08..039C}, microlensing of GWs by PBHs \citep{2019PhRvL.122d1103J}, and constraints from the disruption of ultra-faint dwarfs (UFD) \citep{2016ApJ...824L..31B}. We see that the UDMHs in this scenario are expected to have significant abundance, particularly in the large-mass tail of the mass spectrum, even in the regions already excluded for PBHs by data. The higher number of UDMHs formed during the radiation-dominated era, in comparison to PBHs, is also mentioned in \cite{2009ApJ...707..979R, 2014PhRvD..90h3514K}. In addition, the mass functions in all of the halo formation models turn out to have their peaks at around the same mass ($M\simeq10^5 M_{\odot}$, imposed by the transition scales $k=190~{\rm Mpc}^{-1}$). The DP1 mass function is predicted to have the highest peak, several times larger than the PS prediction. The sudden decrease in the low-mass tail of the mass functions is due to the large-$k$ cutoff in the power spectrum.

Figure \ref{fig_3} illustrates the predicted mass function for the UDMHs in a Universe with the extended Gaussian primordial power spectrum, Eq.\,(\ref{pzetag}). The Gaussian bump enables the exploration of UDMH formation across different mass scales, based on the location of the peak in the power spectrum. As the scale factor increases to $a=a_{\rm eq}$, the differential mass fraction shifts towards higher masses for all mass functions. For the case considered in this work, i.e., with the bump centered at $k_{1}=1.5\times 10^{4}\,\mathrm{Mpc}^{-1}$, the peaks of the UDMH mass functions at $a=a_{\rm eq}$ happen at $M\simeq 10^{3}\,M_{\odot}, 5\times 10^2 M_{\odot}, 2\times10^2 M_{\odot}$, and $10^2 M_{\odot}$ for the DP1, ST, DP2, and PS models, respectively. The results also show significant contributions for the UDMHs in the large $M$-tail, with the DP1 having the highest contribution, again mainly in the regions excluded for PBHs. Similar to the method employed for the first case, we have depicted the mass distribution of PBHs based on the thermal history of the early Universe, while considering the extended Gaussian primordial power spectrum (Eq.\,\ref{pzetag}). In this mass spectrum, the largest contribution can be from solar mass PBHs. There are two additional bumps for PBHs with masses of $10^{-4} M_{\odot}$ and $30 M_{\odot}$, respectively. In this comparison, once again, one can see that for masses greater than $10 M_{\odot}$, a prominent contribution of dark matter can reside in UDMHs arising from DP1, ST, DP2, and PS formalisms, respectively.

Figure \ref{fig_4} indicates the predicted mass function of UDMHs seeded by a narrow Gaussian primordial power spectrum. The results suggest that with the more realistic analytical models for the formation of UDMHs (such as DP1 and ST, which account for physical factors like angular momentum and triaxial collapse), the corresponding dark matter abundance is higher than that predicted by the PS mass function. This abundance demonstrates a time-dependent increase, reaching its maximum near matter-radiation equality. Of particular interest is the remarkable evolution of the dynamical friction factor. Consequently, the DP2 analytical model predicts a negligible abundance of UDMHs at $a=10^{-6}$. However, as time elapses and $a$ approaches $a_{\rm eq}$, this abundance becomes significant. Also, with the assumed primordial power spectrum, a narrow mass spectrum of PBHs can be produced based on the thermal history of the early Universe. This mass spectrum represents a scenario where PBHs account for the entirety of dark matter. This scenario presents two potential avenues for dark matter formation.  PBHs with a mass of approximately $3\,M_{\odot}$ are modeled as primary constituents. Additionally, the clustering of smaller PBHs into UDMHs offers another, potentially significant, source for dark matter. In the MC scenario, i.e., when $f_{\rm PBH}<1$, three observational constraints on the fraction of PBHs are provided in the mentioned mass interval. The constraints are based on the results from microlensing data by EROS, the stochastic background of PBH mergers with LIGO-SGWB, and direct constraints on PBH-PBH mergers with LIGO. In this scenario, it can be seen that the abundance of UDMHs in the dark matter content, while considering more realistic mass functions, falls within the regions that are excluded by observational constraints for PBHs. This means that the abundance of UDMHs, formed by the virialization of dark matter particles in a narrow mass range, can be higher than that of PBHs formed via separate mechanisms.

Figures\,\ref{fig_5} and \ref{fig_6} illustrate the mass functions of smaller-mass UDMHs, seeded by fluctuations described by the power spectra of Eqs.\,(\ref{ppezeta}) and (\ref{pppzeta}) respectively. Given the remarkably high internal density of UDMHs formed on smaller scales, their evolution is expected to be minimally affected by the scale factor. Therefore, for these two cases, we set $a=a_{\rm eq}$ in the calculations for convenience. 

Figure\,\ref{fig_5} illustrates the differential dark matter mass fraction in UDMHs, calculated for the $P^{\rm PE}_{\zeta}(k)$ as the primordial power spectrum with various amplitudes $A_{1}$, corresponding to $P^{\rm PE}_{\zeta}(k_{1})=10^{-1}, 10^{-2}$ and $10^{-3}$ at the characteristic scale $k_1=6\times 10^{6}\,{\rm Mpc^{-1}}$. 
The results are shown for the predictions of the three different halo formation scenarios ST, DP1, and DP2, compared against the PS mass function. The plots show that the DP1 scenario has the largest overall contribution to dark matter compared to ST, DP2, and PS. Also, the predicted mass for the largest halos to form in DP1 is significantly ($\sim$ an order of magnitude) higher than the expected maximum mass in the other three formalisms. We also find that this maximum mass in each model decreases as the amplitude $P^{\rm PE}_{\zeta}(k_{1})$ drops from $10^{-1}$ to $10^{-3}$. This is expected as with lower amplitudes for the perturbations that would seed dense halos, the chances of their formation would decrease. Similar to the previous analysis, the mass spectrum of PBHs is derived from the thermal history of the early Universe within the SC scenario, using the power spectrum of Eq.\,(\ref{ppezeta}). This spectrum results in a peak contribution of PBHs around a mass of $10^{-9} M_{\odot}$. Additionally, in this scenario, UDMHs would form from the clustering of smaller PBHs and contribute significantly to dark matter. In the MC scenario, the observational constraints on the abundance of PBHs are plotted for the extended mass functions \citep{2017PhRvD..96b3514C}. They include the microlensing constraints from the Hyper-Supreme Cam (HSC) \citep{2020PhRvD.102h3021C}, PBH-induced motions in pulsar timing arrays \citep{2019PhRvD.100b3003D}, microlensing constraints from Kepler \citep{2014ApJ...786..158G}, and microlensing constraints from EROS \citep{2007A&A...469..387T}. In this scenario, taking into account more realistic mass functions, the abundance of UDMHs lies within the regions excluded by observational constraints for PBHs.

Similar calculations have been performed for the $P_{\zeta}^{\rm PP}(k)$ for the primordial power spectrum and the results are presented in Figure\,\ref{fig_6}. The mass functions show deviations in the low-mass tail from the predictions for the $P_{\zeta}^{\rm PE}(k)$ spectrum, and are damped with relatively shallower slopes. Apart from this, the overall behavior of the mass functions is approximately similar to the results of Figure\,\ref{fig_5}.

In this work we have focused on adiabatic curvature perturbations. However, it is worth noting that isocurvature perturbations in radiation-dominated era can lead to the formation of UDMHs, in an even more favorable way \citep{1994PhRvD..50..769K}. That is because with adiabatic curvature perturbations, local matter domination is required for UDMHs to form, while PBHs can emerge from radiation fluctuations alone. With isocurvature perturbations, on the other hand, both PBHs and halos would essentially originate from matter perturbations. 

Ultimately, one would expect UDMHs to persist throughout the evolution of the Universe as their internal structure is unlikely to be significantly altered by subsequent accretion of matter \citep{2023MNRAS.518.3509D}. Furthermore, their exceptionally high density, i.e., $\rho \sim 10^{12}\, M_{\odot}/{\rm pc}^{3}$, grants them substantial resistance to phenomena such as tidal stripping within larger structures like galactic halos \citep{2023MNRAS.521.4432S}. It has also been argued that the infrequent UDMH mergers are unlikely to impact the halos' mass fraction and their characteristic internal density values \citep{2019MNRAS.487.1008D, 2019PhRvD.100b3523D}.

\section{Conclusions} \label{sec:iv}
PBHs, originating from substantial density fluctuations, could have a mass distribution spanning several orders of magnitude. In such scenarios, analogous yet less intense, primordial density fluctuations might give rise to an abundance of UDMHs that form alongside PBHs. Having formed early in the radiation-dominated era, these halos are predicted to be highly dense, i.e., about $10^{12}$ times denser than regular cold dark matter halos. Their high density makes them resistant to tidal disruption.

In this work we investigated the formation and evolution of UDMHs in the radiation-dominated epoch. We employed analytical modeling based on the excursion set theory to calculate their abundance. Specifically, we derived the halo mass functions using DP1, DP2 and ST formulations, considering physical factors such as angular momentum, dynamical friction, cosmological constant and triaxial collapse geometry and compared the results with the PS forecasts. The calculations were carried out for four primordial power spectra, with different amplification patterns at specific scales. The power spectra have been chosen to cover subsolar, stellar, and intermediate-mass ranges for UDMHs.

The results demonstrate the higher abundance of UDMHs predicted by the more realistic mass functions considered in this work compared to the PS formalism which highlights the importance of these physical and geometrical factors. In particular the DP1 formulation leads to the highest abundance of UDHMs, as well as predicting the highest maximum mass, for a given primordial power spectrum, compared to the other three formalisms. This implies that angular momentum has a significant impact on halo formation On the other hand, the decrease in the halo abundance for the DP2 mass function suggests the suppression of halo formation due to dynamical friction, although it is still higher than the predictions by the over-simplified PS mass function. The ST mass function, which accounts for triaxial collapse geometry, yields an intermediate abundance between DP1 and DP2 formulations.

In our analysis, we investigated two distinct scenarios concerning the nature of dark matter. The former, SC scenario, proposed a single-component model where dark matter is solely composed of PBHs. Hence, we discussed the probability for UDMHs to have formed via the clustering of smaller PBHs. The results indicate that the abundance of UDMHs, likely formed via the clustering of smaller PBHs and with more realistic mass functions, could significantly account for dark matter content.

Alternatively, the latter, MC scenario, presented a multi-component system where dark matter includes a mixture of particles and PBHs. In this situation, UDMHs would likely form as a result of dark matter particle virialization. This process would be distinct from the formation and clustering of PBHs. To compare the abundance of UDMHs and PBHs as distinct components in this scenario, we also provided observational constraints on the abundance of PBHs. This analysis shows that UDMHs are most likely to be abundant in regions observationally excluded for PBHs. This result is expected to hold, irrespective of the primordial power spectrum, as the power spectra considered in this work cover different scales and have different characteristics. 

Our results highlight the importance of accurate analytical approaches for a better understanding of UDMH formation. This is particularly important as these halos are expected to survive throughout cosmic time, potentially existing as components of present-day dark matter structures since the early Universe. The forecasted increase in their abundance can significantly alter PBH scenario predictions. Future observations have the potential to detect these halos and thereby provide a means to test hypotheses regarding PBH production in the early Universe. Therefore, the prevalence of UDMHs necessitates a re-examination of PBH constraints and predictions across various mass ranges.
\section*{Acknowledgements}
\noindent 
S.F. gratefully acknowledges the Research Council of Shahid Beheshti University for the support under Grant No. 600/1538/2023-11-25.

\bigskip
\bibliography{draft_udmh}
\end{document}